\documentclass[longauth]{aa} 
\usepackage{graphicx}
\usepackage{natbib}
\usepackage{txfonts}
\newcommand{\xirppi}{$\xi(r_p,\pi)$}

\newcommand{\wprp}{$w_p(r_p)$}
\newcommand{\ro}{$r_0$}
\newcommand{\gam}{$\gamma$}

\def\simge{\mathrel{%
   \rlap{\raise 0.511ex \hbox{$>$}}{\lower 0.511ex \hbox{$\sim$}}}}
\def\simle{\mathrel{
   \rlap{\raise 0.511ex \hbox{$<$}}{\lower 0.511ex \hbox{$\sim$}}}}

\DeclareTextSymbol{\degre}{T1}{6}
\DeclareTextSymbol{\degre}{OT1}{23}

\begin{document}
\title{The zCOSMOS Survey. The dependence of clustering on luminosity and stellar mass at $z=0.2-1$ 
  \thanks{Based on observation undertaken at the European Southern Observatory (ESO)
    Very Large Telescope (VLT) under Large Program 175.A-0839.
    Also based on observations with the NASA/ESA Hubble space Telescope,
    obtained at the Space Telescope Science Institute, operated by AURA Inc.,
    under NASA contract NAS 5-26555, with the Subaru Telescope,
    operated by the National Astronomical Observatory of Japan,
    with the telescopes of the National Optical Astronomy Observatory,
    operated by the Association of Universities for Research in Astronomy, Inc. (AURA)
    under cooperative agreement with the National Science Foundation,
    and with the Canada-France-Hawaii Telescope,
    operated by the National Research Council of Canada,
    the Centre Nationla de la Recherche Scientifique de France
    and the University of Hawaii.
  }
}


\author{
       B.~Meneux        \inst{1,2}
  \and L.~Guzzo         \inst{3}
  \and S.~de~la~Torre   \inst{3,4,5}
  \and C.~Porciani      \inst{6,7}
  \and G.~Zamorani      \inst{8}
  \and U.~Abbas         \inst{9}
  \and M.~Bolzonella    \inst{8}
  \and B.~Garilli       \inst{5}
  \and A.~Iovino        \inst{10}
  \and L.~Pozzetti      \inst{8}
  \and E.~Zucca         \inst{8}
  \and S.~Lilly         \inst{7}
  \and O.~Le F\`evre    \inst{4}
  \and J.-P.~Kneib      \inst{4}
  \and C.~M.~Carollo    \inst{7}
  \and T.~Contini       \inst{11}
  \and V.~Mainieri      \inst{12}
  \and A.~Renzini       \inst{13}
  \and M.~Scodeggio     \inst{5}
  \and S.~Bardelli      \inst{8}
  \and A.~Bongiorno     \inst{1}
  \and K.~Caputi        \inst{7}
  \and G.~Coppa         \inst{8,14}
  \and O.~Cucciati      \inst{4}
  \and L.~de~Ravel      \inst{4}
  \and P.~Franzetti     \inst{5}
  \and P.~Kampczyk      \inst{7}
  \and C.~Knobel        \inst{7}
  \and K.~Kova\v{c}     \inst{7}
  \and F.~Lamareille    \inst{11}
  \and J.-F.~Le~Borgne  \inst{11}
  \and V.~Le~Brun       \inst{4}
  \and C.~Maier         \inst{7}
  \and R.~Pell\`o       \inst{11}
  \and Y.~Peng          \inst{7}
  \and E.~Perez~Montero \inst{11}
  \and E.~Ricciardelli  \inst{13}
  \and J.D.~Silverman   \inst{7}
  \and M.~Tanaka        \inst{12}
  \and L.~Tasca         \inst{4,5}
  \and L.~Tresse        \inst{4}
  \and D.~Vergani       \inst{8}
  \and D.~Bottini       \inst{5}
  \and A.~Cappi         \inst{8}
  \and A.~Cimatti       \inst{14}
  \and P.~Cassata       \inst{4}
  \and M.~Fumana        \inst{5}
  \and A.~M.~Koekemoer  \inst{15}
  \and A.~Leauthaud     \inst{16}
  \and D.~Maccagni      \inst{5}
  \and C.~Marinoni      \inst{17}
  \and H.J.~McCracken   \inst{18}
  \and P.~Memeo         \inst{5}
  \and P.~Oesch         \inst{7}
  \and R.~Scaramella    \inst{19}
}

\offprints{B.~Meneux bmeneux@mpe.mpg.de}

\institute{
      Max-Planck-Institut f\"ur Extraterrestrische Physik, Giessenbachstrasse, 85748 Garching-bei-M\"unchen, Germany
\and  Universit\"ats-Sternwarte, Scheinerstrasse 1, Munich D-81679, Germany
\and  INAF-Osservatorio Astronomico di Brera, Via Bianchi 46, I-23807 Merate (LC), Italy
\and  Laboratoire d'Astrophysique de Marseille, UMR 6110 CNRS Universit\'e de Provence, BP8, F-13376 Marseille Cedex 12, France
\and  INAF -- Istituto di Astrofisica Spaziale e Fisica Cosmica, Via Bassini 15, I-20133 Milano, Italy
\and  Argelander Institute for Astronomy, Auf dem H\"ugel 71, D-53121, Germany
\and  Institute of Astronomy, ETH Zurich, Zurich, Switzerland
\and  INAF -- Osservatorio Astronomico di Bologna, Bologna, Italy
\and  INAF -- Osservatorio Astronomico di Torino, Strada Osservatorio 20, I-10025 Pino Torinese (TO), Italy
\and  INAF -- Osservatorio Astronomico di Brera, via Brera 28, Milano, Italy
\and  Laboratoire d'Astrophysique de Toulouse-Tarbes, Universit\'e de Toulouse, CNRS Toulouse, F-31400, France
\and  European Southern Observatory, Garching, Germany
\and  Dipartimento di Astronomia, Universit\`a di Padova, Padova, Italy
\and  Dipartimento di Astronomia, Universit\`a degli Studi di Bologna, Bologna, Italy
\and  Space Telescope Science Institute, 3700 San Martin Drive, Baltimore, MD 21218, USA
\and  Berkeley Lab \& Berkeley Center for Cosmological Physics, University of California, Berkeley, CA 94720, USA
\and  Centre de Physique Theorique, UMR 6207 CNRS Universit\'e de Provence, F-13288 Marseille, France
\and  Institut d'Astrophysique de Paris, Universit\'e Pierre \& Marie Curie, Paris, France
\and  INAF -- Osservatorio Astronomico di Roma, via di Frascatti 33, 00040 Monte Porzio Catone, Italy
}

\date{Received --; accepted --}

\abstract
    {}
    {We study the dependence of galaxy clustering on luminosity 
      and stellar mass at redshifts $z\sim [0.2-1]$, using the first 10K
      redshifts from the zCOSMOS spectroscopic survey of the COSMOS field.
    }
    {We measure the redshift-space correlation functions
      $\xi(r_p,\pi)$ and $\xi(s)$ and the projected function, $w_p(r_p)$ for
      sub-samples covering different luminosity, mass and redshift ranges. 
      We explore and  quantify in detail the observational selection biases
      due to the flux-limited nature of the survey, using ensembles of
      realistic semi-analytic mock samples built from the Millennium
      simulation. We  use the same mock data sets to 
      carefully check our covariance and error estimate techniques,
      comparing the performances of methods based on the scatter in
      the mocks and on bootstrapping schemes.  We finally compare our
      measurements to the cosmological model predictions from the mock
      surveys.
    }
    {At odds with other measurements at similar redshift and in the
      local Universe, we find a weak dependence of galaxy clustering on
      luminosity in all three redshift bins explored.  A mild
      dependence on stellar mass is instead observed, in particular on
      small scales, which becomes particularly evident in the central
      redshift bin ($0.5<z<0.8$), where \wprp\ shows strong excess
      power on scales $>1$~h$^{-1}$ Mpc.  This is reflected in the
      shape of the full $\xi(r_p,\pi)$ that we interpret as produced
      by large-scale structure dominating the survey volume and
      extending preferentially in direction perpendicular to the
      line-of-sight.   Comparing to 
      $z\sim 0$ measurements, we do not see any significant evolution
      with redshift of the amplitude of clustering for bright and/or
      massive galaxies. 
    }
    {This is consistent with previous results and
      the standard picture in which the bias evolves more rapidly for
      the most massive halos, which in turn host the
      highest-stellar-mass galaxies.  At the same time, however, the
      clustering measured in the zCOSMOS 10K data at $0.5<z<1$ for 
      galaxies with $\log(M/M_\odot)\ge 10$ is only marginally
      consistent with the predictions from the mock surveys.  On
      scales larger than 
      $\sim 2$ h$^{-1}$ Mpc, the observed clustering amplitude is
      compatible only with $\sim$~1\% of the mocks. Thus, if 
      the power spectrum of matter is $\Lambda$CDM with standard
      normalization and the bias has no ``unnatural'' scale-dependence,
      this result indicates that COSMOS has
      picked up a particularly rare, $\sim~2-3\sigma$ positive fluctuation
      in a volume of  $\sim 10^6$~h$^{-1}$~Mpc$^3$.  These findings
      underline the need for larger surveys of the $z\sim 1$
      Universe to appropriately characterize the level of structure at
      this epoch. 
    }

    \keywords{Cosmology: observations -- Cosmology: large-scale structure of Universe --
      Surveys -- Galaxies: evolution
    }
      
    \authorrunning{Meneux, B., et al.}
    \titlerunning{Clustering as a function of galaxy luminosity and mass}
      
    \maketitle
%

\section{Introduction}

In the canonical scenario of galaxy formation, galaxies are thought to form
through the cooling of baryonic gas within extended dark matter halos
\citep{whiterees1978}.  The mass of the hosting halo is expected to play a
significant role in the definition
of the visible properties of the 
galaxy, as the total mass in gas and stars, its luminosity, color, 
star formation rate and, possibly, morphology.

Since it is the baryons that form the visible fabric of the Universe,
a major challenge in testing the galaxy formation paradigm is to build
clear connections between these observed properties and those of the
hosting dark-matter halos. 
This is a difficult task, as any direct connection existing initially
between the dark-matter mass and the baryonic component cooling within
the halo is modified by all subsequent dynamical processes affecting
the halo-galaxy system, as merging or dynamical friction.
This is confirmed by simulations, that also show however that galaxy
luminosity and stellar mass do retain in fact memory of the "original"
(not actual) halo mass, i.e. before it experiences a major merger or
is accreted by a larger halo \citep{conroy2006,wang2006,wang2007}.
This gives some hope that measuring the dependence of the galaxy distribution
on galaxy properties one is actually constraining the relationship between
the dark and luminous components of galaxies. 

Measurements of first moments, as the luminosity function or the stellar mass
function, provide a way to understand how these are
related to the total halo mass functions, which can be obtained from
analytic predictions \citep[e.g.][]{press_schechter1974} or n-body
simulations \citep[e.g.][]{warren2006}. Similar investigations can
be made on the second moment, i.e. the two-point correlation function \citep[e.g.][]{springel2006}.
Studies of galaxy clustering in large local surveys have shown how
clustering at $z\sim0$ does depend significantly on several specific
properties.  These include luminosity
\citep{hamilton1988,iovino1993,maurogordato1991,benoist1996,guzzo2000,norberg2001,norberg2002,zehavi2005}, 
color or spectral type \citep{willmer1998,norberg2002,zehavi2002},
morphology \citep{davisgeller1976,giovanelli1986,guzzo1997},
stellar mass \citep{li2006} and environment \citep{abbas2006}.

In recent years it has become possible to extend these
investigations to high redshift, obtaining first indicative
results on how these dependences evolve with time 
\citep{coil2006,pollo2006,phleps2006,meneux2006,daddi2003,meneux2008}. 
The VIMOS-VLT Deep Survey (VVDS) \citep{pollo2006} and the DEEP2
survey \citep{coil2006} in particular, have
provided new insights on the way galaxies of 
different luminosity cluster at $z\sim 1$.  More specifically,
\citet{pollo2006}  
have shown that at these epochs galaxies already show a luminosity segregation,
with more luminous galaxies being more clustered than faint
objects.  At the same time, however, a significant steepening with
luminosity of the shape of their two-point correlation function for
separations $<1-2$~h$^{-1}$~Mpc, is observed.  This 
behaviour is at variance with that at $z\sim 0$. A
similar trend has been observed at the same redshift by the
DEEP2 survey \citep{coil2006}.  Complementarily, \citet{meneux2008}
have shown a positive trend of clustering with stellar mass also at
$z\sim 1$, with a clear evidence for a stronger evolution of the
bias factor for the most massive galaxies \citep[see also][]{brown2008,wake2008}.

The interpretation of the evolution in shape and amplitude of \wprp\
with respect to luminosity and
redshift is particularly interesting in the context of the
halo model for galaxy formation.  In this framework, the observed
shape of $\xi(r)$ (or \wprp) is interpreted as composed by the
sum of two components:
(a) the 1-halo term, which dominates on small scales ($<1-2$~h$^{-1}$~Mpc at the current epoch),
where correlations are dominated by pairs of galaxies living within the same dark-matter halo
(i.e. in a group or cluster);
(b) the 2-halo term on large scales, which is characterized by pairs of galaxies occupying
different dark-matter halos (see \citet{cooraysheth2002} for a review).
\citet{zheng2007} have modelled the luminosity-dependent
\wprp\ from both the DEEP2 (at $z\sim 1$) and SDSS (at $z\sim 0$)
surveys, within such {\sl Halo Occupation Distribution} (HOD) framework. In
this way they establish evolutionary connections between
galaxies and dark-matter halos at these two epochs, providing a self-consistent
scenario in which the growth of the stellar mass depends on the halo
mass.  Similar results are obtained more recently in a combined
analysis of the VVDS-Deep and SDSS data \citep{abbas2009}.

In this paper we use the first 10,000 redshifts from the zCOSMOS
redshift survey (the ``10K sample'') to further explore these
high-redshift trends of clustering with luminosity and mass based on a
new, independent sample. Although shallower than VVDS-Deep and DEEP2
($I_{AB}<22.5$ vs. 24 and 23.5, respectively), zCOSMOS covers a
significantly larger area and samples a volume of $\sim3\times10^6$~h$^{-1}$~Mpc to redshift z=1.2.
This should hopefully help reducing the
effect of cosmic variance (still strong for samples this size
\citep{garilli2008,stringer2009}), while providing a better sampling of the
high-end tail of the luminosity and mass functions.  However,
one main result from this analysis will be 
the explicit demonstration of how strong cosmic variance still is within volumes of
the Universe this size.  The clustering properties of the zCOSMOS
sample in the volume contained within the redshift range $0.4-1$ seem
to lie at the extreme high end of the distribution of fluctuations on
these scales, as it was in fact already suggested by the angular
clustering of the COSMOS data \citep{hjmcc2007}. As we shall see,
these results and those presented in the zCOSMOS series of clustering papers
\citep{porciani2009,delatorre2009,abbas_prep} indicate how cautious one should be
in drawing far-reaching conclusions from the modelling of current
clustering results from deep galaxy surveys. 

A significant part of this paper is dedicated to discussing in detail
these cosmic-variance effects, together with the impact of
incompleteness on the derived results.
This is particularly important when constructing mass-limited sub-samples
from a magnitude-limited survey, which introduces a mass incompleteness
that depends on redshift and stellar mass.
The intrinsic scatter in the galaxy mass-luminosity
relation determines a progressive loss of faint 
galaxies with high mass-to-light ratio. We study in detail the effect of this
incompleteness on the measured clustering both using
the data themselves and mock samples built from
the Millennium simulation.  At the same time, we explore in quite some
detail our ability to characterize measurement errors and the
covariance matrix of our data, comparing estimates from the mock
samples to those from bootstrap resamplings of the data themselves.

The paper is organised as follows. In Sect.~2 and 3 we describe the
zCOSMOS survey and the simulated mock samples used in the analysis,
while in Sect.~4 we describe the selection of luminosity- and
mass-limited subsamples, discussing extensively the 
incompleteness related to this operation; in Sect.~5 we describe our
clustering estimators, while in Sect.~6 we discuss in detail the
observational biases and selection effects, how we account for them
and what is their effect on the measured quantities; in Sect.~7 we
explore and discuss the error budget and how to estimate the
covariance properties of our measurements; in Sect.~8 and 9
we present our measurements of clustering as a function of luminosity
and mass, respectively, while in Sect.~10 we compare these results with
those from other surveys and with simple model predictions; finally,
in Sect.~11 we place these findings in a broader context and discuss
future developments.

Throughout the paper we adopt a cosmology with $\Omega_m = 0.25$,
$\Omega_{\Lambda} = 0.75$. When needed, we also adopt a value
$\sigma_8=0.9$ for the normalization of the matter power spectrum;
this is chosen for consistency with the Millennium simulation, also
used for comparison to model predictions. The Hubble constant is
parameterized via $h=H_0/100$ to ease comparison with previous works.
Stellar masses are quoted in unit of $h=1$.  All length
values are quoted in comoving coordinates. 

\begin{figure}
  \includegraphics[width=9cm]{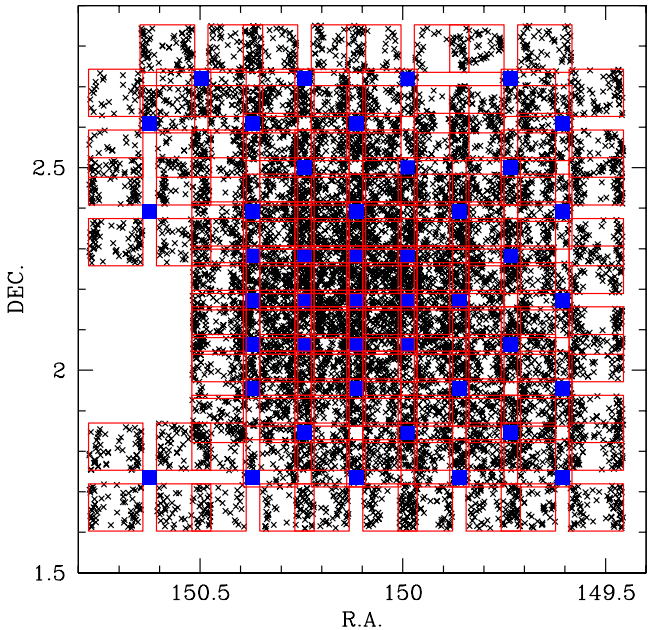}
  \caption{Distribution on the sky of the $\sim 10,000$ galaxies with measured
    redshift (crosses) forming the zCOSMOS  ``10K'' sample.
    The large blue dots
    mark the centres of independent VIMOS pointings, each including
    four quadrants on the sky (as described by the red solid
    lines).
  }
  \label{fig:radec_distr}
\end{figure}

\section{The zCOSMOS survey data}
\label{sec:data}

The zCOSMOS survey \citep{lilly2007} is performed with the VIMOS
multi-object spectrograph at the ESO Very Large Telescope
\citep{lefevre2003_spie}. 600 hours of observation have been allocated
to this program.  These are invested to measure spectra for galaxies
in the COSMOS field \citep{scoville2007a}, targeting: (a) $\sim$20000
galaxies brighter than $I\le 22.5$ 
({\it zCOSMOS Bright}); $\sim$10000 sources at redshift $1.4<z<3.0$
pre-selected using color-color criteria \citep{lilly2007} ({\it zCOSMOS Faint}).
So far, the survey has observed about half of the total ``Bright'' sample,
This is the so called ``10K'' sample used for the
analysis presented in this paper, and is based on the observations of
83 VIMOS pointings over 44 distinct telescope positions on the
sky \citep{lilly2009}. 
These are shown in Fig.~\ref{fig:radec_distr}, where the footprint
of VIMOS (4 quadrants of $\sim 7\times 8$~arcmin$^2$ separated by a
cross about 2~arcmin$^2$ wide) is evident.
About every $3^{rd}$ galaxy has been observed in the field.
The final ``20K'' zCOSMOS sample will be twice larger, reaching
a sampling around 60-70\%.
The correction of the complex angular selection function will be
discussed later in the context of our galaxy clustering measurement. 

Observations are performed using the medium resolution RED grism,
corresponding to $R\sim600$ and covering the spectral range $5550-9650$\AA.
The average error on the redshift measurements has been estimated from
the repeated observations of 632 galaxies,
and is found to be $\sim$100~km~s$^{-1}$ \citep{lilly2009}.
This roughly corresponds to a radial distance error of 1~h$^{-1}$~Mpc.
The reduction of the data to the redshift assignment was carried out
independently at two institutes before a reconciliation process to
solve discrepancies. The quality of each measured redshift was then
quantified via a quality flag that provides us with a 
confidence level (see \citet{lilly2007,lilly2009} for
definition). For the present work, we only use redshifts with flags
1.5 to 4.5 and 9.3 to 9.5, corresponding 
to confidence levels greater than 98\%.  

The zCOSMOS survey benefits of the large multi-wavelength coverage of
the COSMOS field \citep{capak2007}, that with the latest additions now
comprises 30 photometric bands
\citep{ilbert2009} extending well into the infrared. These include
in particular accurate $K$-band and Spitzer-IRAC photometry
over the whole area, which have allowed us to derive relevant physical
properties as rest-frame luminosity and stellar mass with
unprecedented accuracy \citep{zucca2009,bolzonella2009,pozzetti2009}.
\begin{figure*}
  \includegraphics[width=18cm]{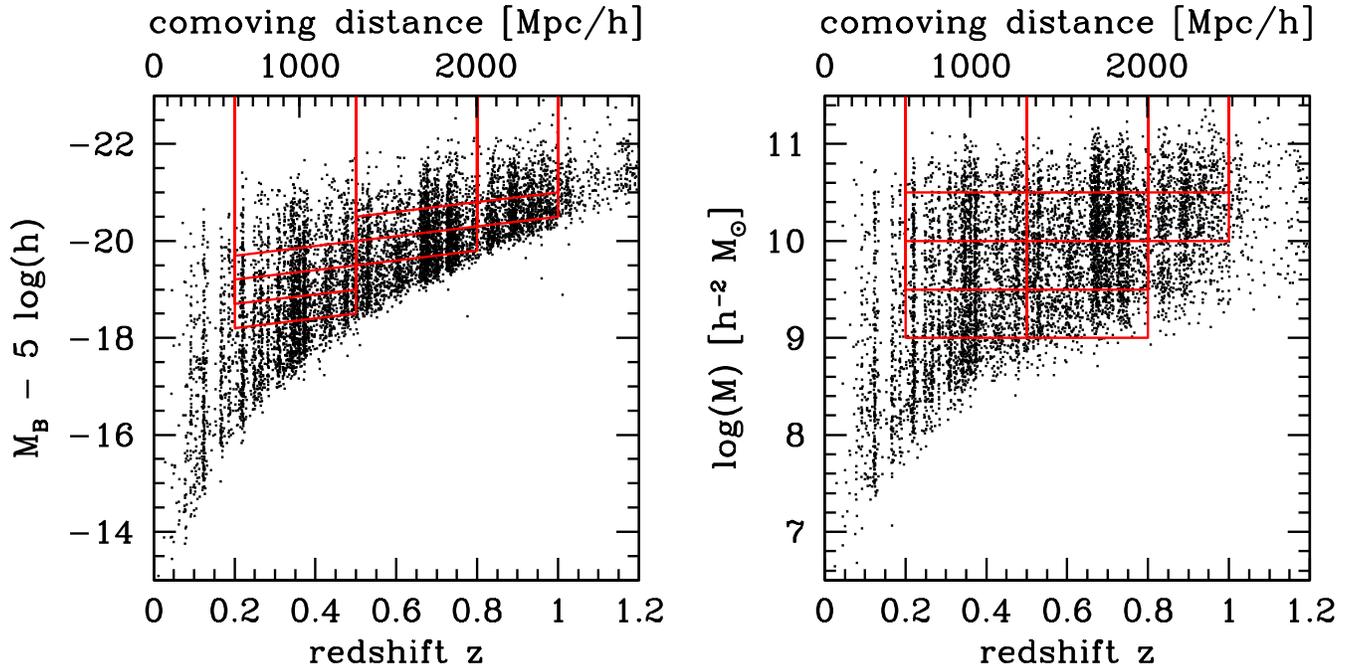}
  \caption{Selection boundaries of the different sub-samples of the
      zCOSMOS 10K survey used in this paper.  {\it Left:}
      luminosity-redshift selection, which accounts for the average
      luminosity evolution of galaxies; {\it Right:} mass-redshift selection.}
      \label{fig:lum_mass_samples}
\end{figure*}

\section{Mock survey catalogues}

In this paper we shall make intense use of mock surveys constructed from the
Millennium simulation \citep{springel2005}. This is done (a) to
understand the effect of our selection criteria on the measured
quantities (Sec.~\ref{sec:xi_incompleteness}); (b) to estimate
the measurement errors and covariance of the data (Sec.~\ref{sec:xierrors}). 

We use two sets of light cones, constructed as explained in 
\citet{kitzbichler2007} and \citet{blaizot2005} by combining dark-matter
halo trees from the Millennium run to the Munich semi-analytical model
of galaxy formation \citep{delucia_blaizot2007}.
The two sets contain 24 $1.4\times 1.4$ deg$^2$ mocks built by 
\citet{kitzbichler2007} and 40 $1\times 1$ deg$^2$ mocks built by
\citet{delucia_blaizot2007}, that we shall name KW24 and DLB40,
respectively.
The main difference between the two sets, in addition to the different survey area,
is that the DLB40 set
contains all galaxies irrespective to any criteria down to the simulation limit
that corresponds roughly to $10^8~M_\odot$, up to redshift $z=1.7$, whereas the KW24 set
contains only galaxies brighter than $I\le22.5$.
This implies that the DLB40 set allows us to select in stellar mass down to very low
masses and test selection effects.
The observing strategy of the zCOSMOS 10K sample was only applyied to the KW24 set,
allowing us to do a carefull error analysis of our measurements.

The Millennium run contains $N=2160^3$ particles
of mass $8.6\times 10^8$~h$^{-1}$~M$_\odot$
in a cubic box of size $500$~h$^{-1}$~Mpc.
The simulation was built with a $\Lambda$CDM cosmological model with
$\Omega_m=0.25$, $\Omega_\Lambda=0.75$, $\sigma_8=0.9$ and
$H_0=73$~km~s$^{-1}$~Mpc$^{-1}$. 
%

\section{Luminosity- and Mass-selected subsamples}

\subsection{Luminosity selection}
\label{sec:lumin_sel}

Absolute magnitudes have been derived for the 10K galaxies using the
code ALF \citep{ilbert2005,zucca2009}, which is based on fitting a Spectral
Energy Distribution to the observed multi-band photometry.
There are various sources of uncertainties to take into account
(errors on apparent magnitudes, number of available photometric bands, method used, \ldots).
A direct comparison with absolute magnitudes derived with the independent
code ZEBRA \citep{feldmann2006} shows consistent estimates with
a small dispersion of $\sigma\sim0.05$ magnitudes, in particular in the B band.
This can resonably be considered as the typical error on our absolute magnitudes.
 
For our analysis, the goal is to define luminosity-limited
samples that are as close as possible to truly volume-limited samples, i.e.
with a constant number density.  This should be possibly done within a
few independent redshift ranges.
The size of the redshift slices in which to split the sample has to be
chosen as a
compromise between two aspects: (a) we want it to be large enough as
to have sufficient statistics and provide a good measurement of
clustering; however, (b) we do not want it to be too large, as to
avoid significant evolution within each redshift bin.

However, we know that all through the overall redshift range covered
by the zCOSMOS survey ($0.2<z<1.1$) the luminosity of galaxies
evolves, with a clear change in the characteristic parameters of the
luminosity function \citep{ilbert2005}.
This evolution does depend on the morphological/spectral type of the galaxy considered.
To be able to select a nearly volume-limited sample within
a given redshift interval, we need to
take the corresponding evolution into account. This can be
realistically done only in a statistical way, looking at the
population-averaged evolution of the global luminosity function.  

We have therefore considered the observed luminosity function measured
from the same data \citep{zucca2009} and modelled its change with redshift as a
pure luminosity evolution (i.e. keeping a constant slope $\alpha$ and 
normalisation factor $\Phi^*$), which is a fair description of the
observed behaviour. We find that the characteristic absolute
magnitude $M^*(z)$ evolves with redshift as 
\begin{equation}
M^*(z) = M^*_0 + A z \,\,\,\, ,
\end{equation}
where $A\sim -1$. 
In the companion paper, \citet{delatorre2009} split the
zCOSMOS galaxy samples in 3 morphological classes. They observe
different luminosity evolutions for elliptical, spiral and irregular
galaxies, with $A$ varying from $\sim-0.7$ for to $\sim-1.2$ but
with large uncertainties making $A=-1$ compatible for all
classes. \citet{porciani2009} reach similar conclusions when
dividing the zCOSMOS 10K sample into 3 color classes.

We therefore define our luminosity-limited samples by
an effective absolute magnitude cut at $z=0$, $M_{B,cut}$ and
including all galaxies with $M_B(z) - 5 log(h) \le\ M_{B,cut}-z$.  The
resulting selection loci for different values of $M_{B,cut}$ are
plotted over the data in the luminosity-redshift plane in the left
panel of Fig.~\ref{fig:lum_mass_samples}.  As evident from the figure,
the faintest allowed threshold $M_{B,cut}$ depends on the
redshift range considered, i.e. z=[0.2-0.5], z=[0.5-0.8] and
z=[0.8-1.0].  The details of the resulting samples are described in
Table~\ref{tab:lum}.
\begin{table}
  \caption{Properties of the luminosity-selected samples}
  \label{tab:lum}
  \centering

  \begin{tabular}{lcccc}
    \hline\hline
Sample   & Redshift & Mean     & $M_{B,cut}$ &  Number of \\
         & range    & redshift &   (z=0)    &  galaxies  \\
    \hline
L1.1     & 0.2-0.5  &  0.37    &   -18.00   &   1892     \\
L1.2     & 0.2-0.5  &  0.37    &   -18.50   &   1311     \\
L1.3     & 0.2-0.5  &  0.37    &   -19.00   &    811     \\
L1.4     & 0.2-0.5  &  0.37    &   -19.50   &    469     \\
    \hline
L2.1     & 0.5-0.8  &  0.67    &   -19.00   &   1848     \\
L2.2     & 0.5-0.8  &  0.67    &   -19.50   &   1025     \\
L2.3     & 0.5-0.8  &  0.67    &   -20.00   &    441     \\
    \hline
L3.1     & 0.8-1.0  &  0.91    &   -19.50   &    971     \\
L3.2     & 0.8-1.0  &  0.91    &   -20.00   &    447     \\
    \hline
  \end{tabular}
\end{table}

\begin{figure*}
  \includegraphics[width=18cm]{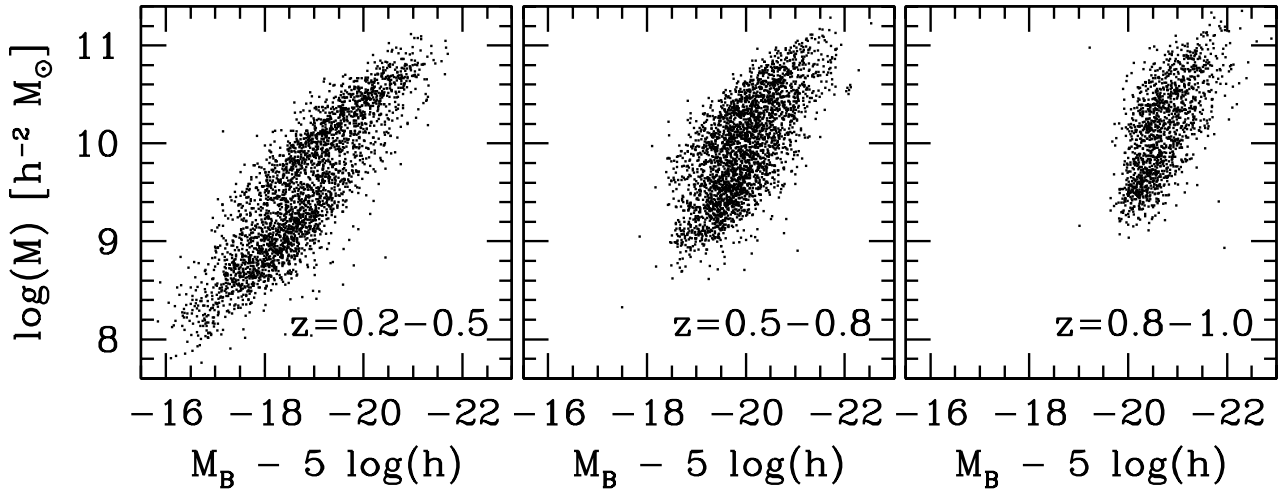}
  \caption{The observed relationship between stellar mass and
    luminosity for galaxies in the 10K sample, within the three
    redshift ranges studied in this paper. The left panel shows
    an aspect of the galaxy bimodality, with red galaxies more massive
    and brighter than blue ones.
  }  
  \label{fig:mass_lum_relation}
\end{figure*}

\subsection{Mass selection}
\label{sec:datamass}

%

\begin{table}
  \caption{Properties of the mass-selected samples} 
  \label{tab:mass}
  \centering
  \begin{tabular}{lccccc}
    \hline\hline
Sample   & Redshift & Mean     &\multicolumn{2}{c}{ $log(M/M_\odot)$ }              & Number of \\
         & range    & redshift &           range       &          median           & galaxies   \\
    \hline
M1.1     & 0.2-0.5  &  0.36    &   $\ge \ \  9.0$      &           9.80             &  2159    \\
M1.2     & 0.2-0.5  &  0.37    &   $\ge \ \  9.5$      &          10.09             &  1445    \\
M1.3     & 0.2-0.5  &  0.36    &   $\ge 10.0$          &          10.36             &   827    \\
M1.4     & 0.2-0.5  &  0.37    &   $\ge 10.5$          &          10.66             &   275    \\
    \hline
M2.1     & 0.5-0.8  &  0.66    &   $\ge \ \  9.0$      &           9.97             &  2831    \\
M2.2     & 0.5-0.8  &  0.66    &   $\ge \ \  9.5$      &          10.12             &  2276    \\
M2.3     & 0.5-0.8  &  0.67    &   $\ge 10.0$          &          10.38             &  1366    \\
M2.4     & 0.5-0.8  &  0.67    &   $\ge 10.5$          &          10.68             &   477    \\
    \hline
M3.1     & 0.8-1.0  &  0.90    &   $\ge 10.0$          &          10.46             &   755    \\
M3.2     & 0.8-1.0  &  0.90    &   $\ge 10.5$          &          10.73             &   344    \\
    \hline
  \end{tabular}
\end{table}

Stellar mass has become a quantity routinely measured 
in recent years, thanks to surveys with multi-wavelength photometry,
extending to the near-infrared \citep[e.g.][]{rettura2006}, although
some uncertainties related to the detailed modelling of stellar
evolution remain \citep{pozzetti2007}. This has made studies of
clustering as a function of stellar mass possible for large
statistical samples. We used stellar masses estimated by fitting the
Spectral Energy Distribution (SED), as sampled by the large multi-band
photometry, with a library of stellar population models based on
\citet{bc2003}.
We used the code {\it Hyperzmass}, a modified version
of the photometric redshift code {\it Hyperz} \citep{bolzonella2000}.
The typical error on stellar masses is $\sim$0.2~dex.
The method and accuracy of these measurements are
fully described in \citet{bolzonella2009} and \citet{pozzetti2009}.

We have thus constructed a set of mass-selected
samples, containing galaxies more massive than a given
threshold. We choose the same redshifts ranges as used for the
luminosity selected samples. The  properties of the selected
subsamples are summarized in
Table~\ref{tab:mass} and represented in
Fig.~\ref{fig:lum_mass_samples}. 

\subsection{Mass completeness}
\label{sec:mass-incompleteness}

\begin{figure}
  \includegraphics[width=9cm]{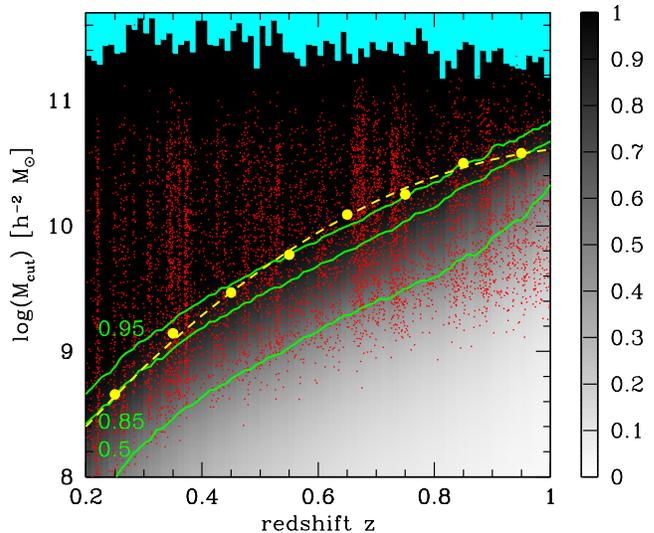}
  \caption{Estimate of how the completeness in stellar mass changes
    as a function of redshift, due to the survey flux limit
    ($I_{AB}<22.5$). The shaded grey area and green contours describe
    the loci of constant completeness. They are derived from the DLB40 mock
    samples of $1\times1$ deg$^2$ and defined as the fraction of
    observed ($I_{AB}\le 22.5$) galaxies over the total number in a
    given cell with size $\Delta z=0.01$ and
    $\log(M)\ge\log(M_{cut})$.
    The red points superimposed correspond to the actual data of the 10K
    sample.
    The yellow points and dotted line show the 95\% M/L ratio
    completeness level derived independently by \citet{pozzetti2009} (see text),
    directly from the observed data.
    The agreement between the two estimates,
    from the data and from the Millennium mocks,
    is remarkable and adds confidence in the use of the simulated samples.
  } 
  \label{fig:mass_completeness_map}
\end{figure}
Due to the flux-limited nature of surveys like zCOSMOS  ($I_{AB}<22.5$)
the lowest-mass samples are affected to varying degree by incompleteness
related to the scatter in the mass-luminosity relation (Fig.~\ref{fig:mass_lum_relation}). 
This introduces a bias against objects
which would be massive enough to enter the mass selected samples, but
too faint to fulfill the apparent magnitude limit of the survey. 
These missed high mass-to-light ratio galaxies will be those
dominated by low-luminosity stars, i.e. the red and faint objects.
Clearly, if this is not accounted for in
some way, it would inevitably affect 
the estimated clustering properties, with respect to a truly complete,
mass-selected sample \citep{meneux2008}.
It is therefore necessary to understand in detail
the effective completeness level in stellar mass of the samples that
we have defined for our analysis.

\citet{meneux2008} have used 2 different methods to explore and
quantify the completeness limit in stellar mass as a function of
redshift.  The first is based on the observed scatter in the
mass-luminosity relation, obtained from the data themselves and
extrapolated to fainter fluxes.
The second instead makes use of mock survey samples, under the hypothesis
that they provide a realistic description of the mass-luminosity
relation and its scatter:
the DLB40 set of mock survey catalogues that are complete
in stellar mass are ``observed'' under the same conditions as the
real data, i.e., selected at $I\le22.5$.
The completeness is then simply defined, for a given redshift
range and mass threshold, as the ratio of the number of galaxies
brighter than the zCOSMOS flux limit over those at any flux.
Interestingly, even if  
this method is model-dependent (in particular, on the prescription of
galaxy formation used in the semi-analytical models), this
approach leads to similar completeness limits than the first one.  
The results of this second exercise are shown, as a function of
redshift and mass threshold and for a flux limit $I\le 22.5$, in
Fig.~\ref{fig:mass_completeness_map}.
Completeness is estimated in narrow redshift ranges ($\Delta z=0.01$) for
different mass thresholds $M_{cut}$ increasing from $10^8$ to
$10^{11.7}~h^{-2}~M_\odot$ with a step of $10^{0.01}~h^{-2}~M_\odot$.
A large fraction of low-mass objects is clearly missed at high
redshift.

The yellow points and dashed line in Fig.~\ref{fig:mass_completeness_map}
have been estimated from the observed scatter in the M/L
relation of the data, and are defined at each redshift as the lower
boundary, $M_{min}(z)$, including above it 95\% of the mass distribution
\citep{pozzetti2009}.

It is very encouraging to notice the very good agreement between
this independent estimation from the data and that
based on the DLB40 set of mock catalogues.
Table~\ref{tab:masscomp} summarises the completeness estimates
derived from these mock catalogues for each of the 10 zCOSMOS
galaxy samples defined in Table~\ref{tab:mass}. The sample M2.1 shows
the strongest incompleteness: 65.1\% of the galaxies more massive than
$10^9~h^{-2}~M_\odot$ are fainter than $I=22.5$ at z=[0.5-0.8] and
then, not included in our sample.
In Sec.~\ref{sec:xi_incompleteness} we shall discuss the effects
of this incompleteness on the galaxy clustering measurement.

\begin{table}
  \caption{The completeness in stellar mass of mass-selected mock
    sub-samples reproducing the properties and selection criteria of our 10K
    data samples. Completeness is defined as the percentage of all
    galaxies above the mass limit which are actually included in the
    sample. 
} 
  \label{tab:masscomp}
  \centering
  \begin{tabular}{lccc}
    \hline\hline
Sample   & Redshift & Stellar mass  ($log(M/M_\odot)$) & Completeness\\
         & range    &             range               &  \\
 \hline
M1.1     & 0.2-0.5 & $\ge$ 9.0 & 0.783 \\
M1.2     & 0.2-0.5 & $\ge$ 9.5 & 0.972 \\
M1.3     & 0.2-0.5 & $\ge$10.0 & 1.000 \\
M1.4     & 0.2-0.5 & $\ge$10.5 & 1.000 \\
 \hline
M2.1     & 0.5-0.8 & $\ge$ 9.0 & 0.349 \\
M2.2     & 0.5-0.8 & $\ge$ 9.5 & 0.652 \\
M2.3     & 0.5-0.8 & $\ge$10.0 & 0.919 \\
M2.4     & 0.5-0.8 & $\ge$10.5 & 0.996 \\
 \hline
M3.1     & 0.8-1.0 & $\ge$10.0 & 0.571 \\
M3.2     & 0.8-1.0 & $\ge$10.5 & 0.882 \\
    \hline
  \end{tabular}
\end{table}

\section{Estimating the two-point correlation function}
\label{sec:xi_technic}

The two-point correlation function is
the simplest estimator used to quantify galaxy clustering, being
related to the second moment of the galaxy distribution, i.e. its
variance.  In practice, it describes the excess probability $\xi(r)$ to
observe a pair of galaxies at a given separation $r$, with respect to that
of a random distribution \citep{peebles1980}. Here we shall estimate the
redshift-space correlation function \xirppi, which allows one to
account and correct for the effect of peculiar motions on the pure
Hubble recession velocity. In this case, galaxy
separations are split into the tangential and radial components,
$r_p$ and $\pi$ \citep{dp83,fisher1994}.

The real-space correlation function $\xi_R(r)$ can be recovered by
projecting \xirppi\ along the line-of-sight, as
\begin{equation}
  w_p(r_p) \equiv 2 \int_0^\infty \xi(r_p,\pi) d\pi = 2 \int_0^\infty
  \xi_R\left[(r_p^2+y^2)^{1/2}\right] dy
  \label{wpdef}
\end{equation}
\begin{figure}
  \includegraphics[width=9cm]{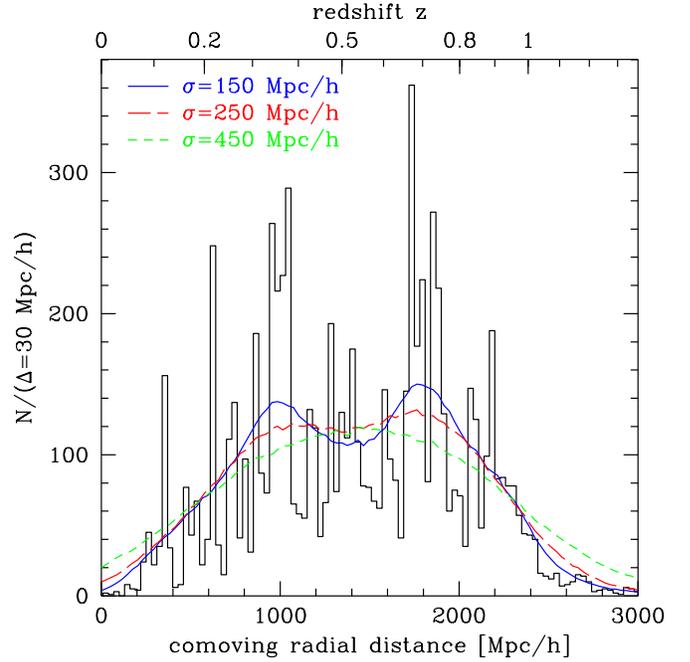}
  \caption{Overall radial distribution of the zCOSMOS 10K sample,
    compared to three different smoothed distributions. These are
    obtained by filtering the observed data with a Gaussian kernel of
    increasing $\sigma$=150, 250 and 450~h$^{-1}$~Mpc. The first two smoothed 
    curves retain information of the two large structures located at
    $\sim$1000 and $\sim$1800~h$^{-1}$~Mpc along the line-of-sight,
    while the third one overestimates the number density of galaxies 
    at low and high redshift.}
  \label{fig:nz_all}
\end{figure}
\begin{figure}
  \includegraphics[width=9cm]{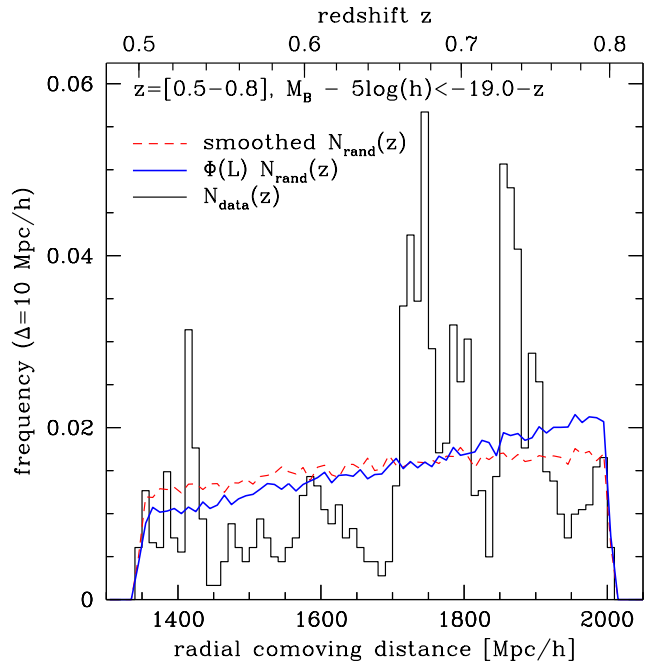}
  \caption{The radial distribution of the luminosity selected sample
    L2.1 is compared to a smoothed curved (with a kernel of
    $\sigma$=450~h$^{-1}$~Mpc - dashed curve) and  a radial
    distribution generated from the integration of the luminosity
    function (solid curve). The latter is consistent with that
    expected for such galaxy sample.} 
      \label{fig:random_samp_radial_distrib}
\end{figure}
For a power-law correlation function, $\xi_R(r)=(r/r_0)^{-\gamma}$,
this integral can be solved analytically and fitted to the observed
\wprp\ to find the best-fitting values of the correlation length
\ro\ and slope \gam\ \citep[e.g.,][]{dp83}.  In computing \wprp,
a finite upper integration limit has to be chosen in practice. Its
value has to be large enough as to include most of the clustering
signal dispersed along the line of sight by peculiar motion.
However, it must not be too large, to avoid adding just noise,
which is dominant above a certain $\pi$.
Previous works \citep{pollo2005} have 
shown that, for similar data, the best results are obtained with
an integration limit $\pi_{max}$ between 20 and 40~h$^{-1}$~Mpc.
Our tests show that the scatter in the recovered \wprp\ is obtained
using the lowest value in this range.
This can introduce a 5-10\% under-estimate in the recovered
large-scale amplitude, which can be accounted for when
fitting a model to \wprp.   In the following, we shall in general use
$\pi_{max}=20$~h$^{-1}$Mpc and show examples of how the amplitude is
biased by this choice for the real data.

To estimate \xirppi\ from each galaxy sample, we use
the standard estimator of \citet{lansal1993}:
\begin{equation}
  \xi(r_p,\pi) = \frac{N_R(N_R-1)}{N_G(N_G-1)} \frac{GG(r_p,\pi)}{RR(r_p,\pi)}
  - \frac{N_R-1}{N_G} \frac{GR(r_p,\pi)}{RR(r_p,\pi)} + 1
  \label{lseq}
\end{equation}
where $N_G$ is the mean galaxy density (or, equivalently, the total
number of objects) in the sample; $N_R$ is the mean density of a
catalogue of random points distributed within the same survey volume and with
the same selection function as the data;
$GG(r_p,\pi)$ is the number of independent galaxy-galaxy pairs with
separation between $r_p$ and $r_p+dr_p$ and between $\pi$ and $\pi+d\pi$;
$RR(r_p,\pi)$ is the number of independent random-random pairs
within the same interval of separations and $GR(r_p,\pi)$ represents
the number of galaxy-random cross pairs.

\section{Observational biases and selection effects}
\subsection{Correction of VIMOS angular foot-print and varying sampling}

To properly estimate the correlation function from the 10K zCOSMOS data, we need
to correct for its spatial sampling rate, which is on average $\sim$30\%,
but varies with the position on the sky due to the VIMOS foot-print
and the superposition of multiple passes (see Fig.~\ref{fig:radec_distr}).
The correction scheme used here is an evolution of that discussed in
\citet{pollo2005}, but with a simplified weighting scheme. The main
differences of this sample with respect to the VVDS-Deep data used by
\citet{pollo2005}, are that:
(a) this sample is 1.5 magnitudes brighter,
(b) the spectra are taken with higher resolution, which produces
longer spectra and thus less objects along the dispersion direction and,
(c) there are as many as 8 repeated observations (``passes'') covering each point
on the sky in the central area of the COSMOS field. 
The net result of these differences can be appreciated in Fig.~\ref{fig:radec_distr}:
the sample is characterized by a well-sampled central region, but also
by rather sparsely sampled VIMOS pointings in the outskirts of the
field.  In particular, these external pointings clearly show target
galaxies concentrated along rows. This effect is produced by the
significant length of the spectra on the CCD in medium-resolution
mode: not more than 2 spectra can be aligned on top of each other on
the detector in each quadrant, which results into the observed two
``stripes''.  This is significantly different from what happens in the
low-resolution observing mode \citep[as e.g. in VVDS-Deep,][]{lefevre2005_vvds},
where spectra are shorter and up to four of
them can be packed along the same column on the CCD. 

We have tested three different algorithms to correct for the angular
selection function of the survey, obtaining comparable results.
Other weighting schemes use in particular the angular correlation
functions of the 10K sample and the photometric catalogue to correct
for the non uniform spatial sampling rate.
These methods are discussed in the parallel clustering
analyses by \citet{delatorre2009} and \citet{porciani2009}.
In the latter paper in particular, comparative tests of the three
algorithms are presented. 

Since the sub-samples analyzed in this work are essentially
volume-limited (above the luminosity/mass completeness limits),
we do not need to apply any further minimum-variance
weighting scheme \citep[as e.g. the $J_3$ weighting,][]{fisher1994}.
This is normally necessary for purely flux-limited surveys in 
which the selection function varies significantly as a function of
redshift, such that different parts of the volume are sampled by
galaxies with different luminosities and number densities
\citep[e.g.,][]{li2006}.  

\subsection{Construction of reference random samples}
 
A significant source of uncertainty that we encountered in estimating
two-point functions from our 10K sub-samples is related to the
construction of the random sample and in particular to its redshift
distribution.  We soon realized that the strongly clustered nature of
the COSMOS field along the line of sight, with several dominating
structures at different redshifts, required some particular care
as not to generate systematic biases in the random
sample.  These superclusters are already evident as vertical stripes
in Fig.~\ref{fig:lum_mass_samples} and even more clearly in the
redshift histogram of Fig.~\ref{fig:nz_all}.  Note the big
``walls'' at $z=0.35$, 0.75 and 0.9, which are also clearly identified
by the density field reconstruction of \citet{kovac2009}. 

A standard way to generate a random redshift coordinate 
accounting for the radial selection function of the data uses a
Gaussian-filtered version of the data themselves. This is normally
obtained using smoothing kernels with a dispersion $\sigma$ (in
co-moving coordinates) in the range 150$-$250~h$^{-1}$~Mpc.  The
results of applying this technique to the current 10K data are shown
in Fig.~\ref{fig:nz_all}.  One notes how for smoothing scales of 150
and 250~h$^{-1}$~Mpc the curves still retain memory of the two
largest galaxy fluctuations.  These are erased only when a very strong
smoothing filter (450~h$^{-1}$~Mpc) is adopted.  However, in this case
the smoothed curve is unable to follow correctly the global shape of
the distribution, over-estimating the number density in the lowest and
highest redshift ranges. The situation for our specific analysis,
however, is somewhat simpler than this general case.
Our luminosity-limited or mass-limited samples are in principle
``volume-limited'', i.e. samples that -- if properly selected -- should
have a constant density within the specific redshift bin. One such case is
shown in the zoom of Fig.~\ref{fig:random_samp_radial_distrib}, where
the redshift distribution in the range $z=[0.5,0.8]$ is plotted. 

An alternative way to generate the radial distribution of the random
sample is to integrate the galaxy luminosity function (LF) in steps along the
redshift direction, computing at each step a value for the density of
galaxies expected at that redshifts.  Ideally, the LF
can be measured from the sample itself and would include any detected
evolution of its parameters.  This is what we did here, using the
evolving LF parameters presented in the companion dedicated paper (Zucca et
al. 2009).  The red dashed line in
Fig.~\ref{fig:random_samp_radial_distrib} shows the result one obtains
if smoothing with a kernel of 450~h$^{-1}$~Mpc (dashed), compared to that
obtained from the integration of the LF (solid). The latter is fully
consistent with that expected from a truly volume-limited sample with
the given selection criteria,
with the number of objects increasing as the square of the radial comoving distance.

\subsection{Effect of mass incompleteness on $w_p(r_p)$}
\label{sec:xi_incompleteness}

As discussed in Sec.~\ref{sec:mass-incompleteness} when we constructed
our mass-limited samples, a fraction of galaxies more massive 
than the formal mass threshold are in fact lost due to the limiting
$I_{AB}<22.5$ flux cut of the survey. This becomes more and more
important with increasing redshift. As we said this population
of missing galaxies is inevitably dominated by red objects with high
mass-to-light ratio \citep{meneux2008}, which are known to 
cluster more strongly than the average population
\citep{meneux2006,coil2008,hjmcc2008}.
Having defined our clustering tools, we can now further extend the
analysis of Sec.~\ref{sec:mass-incompleteness} and use the DLB40 mock
samples to quantify directly the effect this has on the measured
\wprp.   We thus computed the statistics for each of the mocks, which
are complete down to very small masses ($\sim10^8~M_\odot$), with
and without applying the apparent-magnitude cut.  Clearly, we are
making here a very strong hypothesis, i.e. that the simulated samples
have intrinsic clustering properties (and their relation to the
galaxy's $M/L$ ratio), that are similar to those of real data.

\begin{figure}
  \includegraphics[width=9cm]{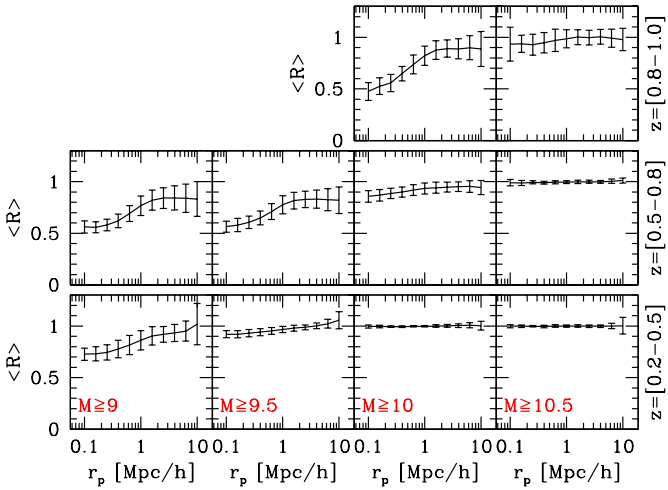}
  \caption{The effect of stellar mass incompleteness 
    on the measured $w_p(r_p)$, estimated from the $1\times
    1$ deg$^2$ Millennium mock samples.  The figure shows the average
    of the quantity $R$ over 40 mock samples as a function of $r_p$.
    $R$ is defined as
    the ratio of the estimates of \wprp\ with and without the flux cut
    at $I_{AB}=22.5$, i.e. for a sample mimicking the 10K selection
    and a sample 100\% complete in stellar mass.
  }
  \label{fig:mass_completeness_clustering}
\end{figure}

The ratio of these two estimates (``true'' over ``observed'') averaged over
the 40 mock catalogues is shown on Fig.~\ref{fig:mass_completeness_clustering}.
For a mass selection which is 100\% complete within the given redshift bin
we would measure $\left<R\right>=1$ at all separations.
We can see that the only mass range for which this is strictly happening
at any redshift is that with $\log(M/M_\odot)\ge $10.5.
For smaller mass samples we see a clear reduction of the clustering amplitude.  
However, we can also see that for most samples the shape of \wprp\ is distorted
mainly only below $< 1$~h$^{-1}$~Mpc.  Above this scale, the mass
incompleteness introduces an amplitude reduction up to $\sim 20\%$ in the
worst cases.  This will have to be considered when comparing our
measurements with models (although keeping in mind that these
estimates come from simulated data, not from real observations). For
general comparisons, however, the amount of amplitude reduction of
\wprp\ is typically negligible on scales larger than $\sim$1~h$^{-1}$~Mpc,
given the statistical errors of the data measurements. 

\section{Systematic and statistical errors on correlation estimates}
\label{sec:xierrors}

\begin{figure}
  \includegraphics[width=9cm]{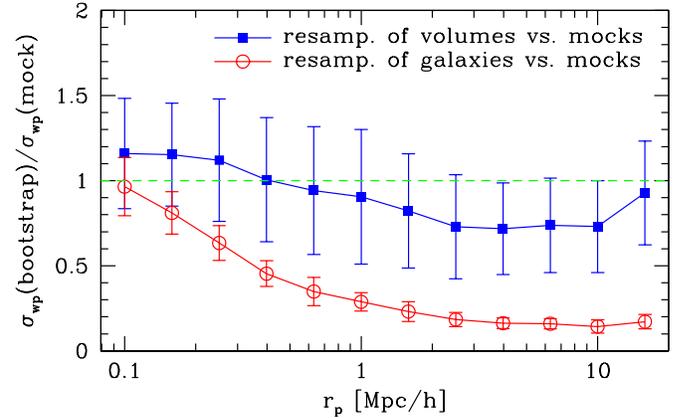}
  \caption{Ratio of the diagonal errors on  $w_p(r_p)$,
    obtained through the {\it bootstrap} resampling
    method to the ``true'' ones obtained from the variance of 24 mock
    catalogues. 
    Filled and open symboles correspond to two different bootstrapping
    techniques, resampling respectively sub-volumes of the survey or
    single galaxies.  The former technique clearly provides a standard
    deviation which is closer to the ``true'' one obtained from
    repeated measurements.} 
  \label{fig:sigmeanbootmock}
\end{figure}

The derivation of realistic errors on the galaxy correlation function
has been the subject of debate since its early measurements
\citep[see e.g.,][]{bernstein1994}. In particular, it is well-known that the
measured values of the two-point correlation function on different
scales are not independent.  This means that, e.g., the bins of \wprp\
have a degree of correlation among them, that needs to be taken into
account when fitting a model to the observed values.  This can be done
if we are able to reconstruct the $N\times N$ covariance (or
correlation) matrix of the $N$ bins \citep{fisher1994}.  

In a recent paper, \citet{norberg2008} compare in
detail three different methods for estimating the covariance matrix of
a given set of measurements.  These use: (a) the ensemble
variance from a set of mock catalogues reproducing as accurately as
possible the clustering properties and selection function of the real
data; (b) a set of {\it bootstrap} resamplings of  
the volume containing the data, and (c) a so-called {\it jack-knife}
sub-set of volumes of the survey.  In this latter case, the survey
volume is divided into $N_V$ sub-volumes and the statistics under study
is re-computed each time excluding one of the sub-parts.
In the ``block-wise'' incarnation of the bootstrap technique
\citep[][method ``b'']{porciani2002}, instead,
$N$ sub-volumes are selected each time {\it with repetition}, i.e.
excluding some of them, but counting two or more times some others as
to always get a global sample with the same total volume.  We note
however that historically there are two possible ways to resample
internally the data set.  The classical ``old'' bootstrap
\citep{ling1986} entailed boot-strapping the sample
``galaxy-by-galaxy''.  This means picking randomly each time a sample
of $N_G$ galaxies among our data set of $N_G$ galaxies, allowing
repetitions.  In this way, within one bootstrap realization a
galaxy can be selected more than once, while some others are never
selected.  This technique has been shown to lead in general to some
under-estimation of the diagonal errors 
\citep{fisher1994}. Here we shall test directly also this aspect.

The advantage of using mock samples is that, under the assumption that
these are a realistic realization of the real data, they allow us to
obtain a true ensemble average and standard deviation from samples with the
same size as the data sample, including both Poissonian noise and
cosmic variance.  Unfortunately, the covariance properties derived
from mock samples are not necessarily a good description of those of
the real data, thus making the use of the derived covariance matrix
(e.g. in model fitting) doubtful.  Conversely, depending on the sample
size, {\it jack-knife} or volume-bootstrap covariance matrices can exacerbate
peculiarities of some sub-regions, again not representing adequately
the true covariance properties of the data. 

For the present investigation, we
spent considerable effort to understand how to best
estimate a sensible covariance matrix for our $w_p(r_p)$ measurements.
The available mock samples were crucial as to allow us to perform
direct comparisons of the performances of the different techniques.
After some initial attempts, we excluded the {\it jack-knife} method
because of the limited size of the survey volume.  We then performed a
direct comparison of the covariance matrices derived through the
bootstrap technique and from the KW24 mock catalogues.  For the
bootstrap method, we decided to test directly how galaxy- and volume-bootstrap
were performing.  We concentrated on the redshift range z=[0.5-0.8]
selecting simulated galaxies brighter than $M_B-5log(h)\le -19.5-z$.
After computing the correlation function $w_p(r_p)$ for all 24 mock
samples, we constructed for each of them: (a) 100 galaxy-galaxy
bootstrap samples and (b) 100 volume-volume bootstrap samples.
In the latter case, we considered 8 equal sub-volumes, defined as
redshift slices within the redshift range considered. The number of
sub-volumes was choosen as the best compromise between having enough
of them and not having too small volumes. With this choice, their
volume is $\sim1.4\times10^5$~h$^{-3}$~Mpc$^3$ for the samples with
z=[0.5-0.8] and z=[0.8-1.0] and $\sim0.6\times10^5$~h$^{-3}$~Mpc$^3$ 
for z=[0.2-0.5]. The two bootstrap techniques lead to a total of
4800 samples and corresponding estimates of 
$w_p(r_p)$.  We then calculated the covariance (and correlation)
matrices for each of these two cases along with the one derived from
the correlation function of the 24 mocks themselves. 

In Fig.~\ref{fig:sigmeanbootmock}  we show a comparison of the
standard deviations derived from the two bootstrap techniques, to that
derived from the 24 mocks.  In each case, these values correspond by
definition to the square root of the diagonal elements of
the covariance matrix.  In the plot we show the mean (over the 24
mocks) of the ratio 
of $\sigma_{w_p}$ from the bootstrap to the ``true'' one from the 
ensemble of mock surveys.  This shows clearly how the {\it r.m.s.} values
obtained with the single-galaxy bootstrap grossly underestimate the
true variance, up to one order of magnitude on large scales.
Bootstrapping by volumes produces a better result, providing a
realistic estimate of $\sigma_{w_p}$ between 0.1 and 1~h$^{-1}$~Mpc, and
a 20-25\% under-estimate on larger scales.

\begin{table}
  \caption{The 5 main eigenvalules of the correlation matrix derived
    with the bootstrap resampling, respectively of galaxies (first column)
    and sub-volumes (col.~2),
    and from the ensemble variance of the 24 mocks (col.~3).
    For the two latter cases, each
    mock sample is used in turn as ``data'' and the reported eigenvalues
    are the obtained as the average over the 24 mocks.
}
  \label{tab:eigencorr}
  \centering
  \begin{tabular}{lccc}
    \hline\hline
  eigenvalue &\multicolumn{2}{c}{bootstrap} &    mocks   \\
             &    galaxies  &     volumes   & catalogues \\
\hline
$\lambda_1$  &     3.87784  &     8.28193   &   11.82062 \\
$\lambda_2$  &     1.95834  &     2.17866   &  \ 0.15708 \\
$\lambda_3$  &     1.36841  &     0.71635   &  \ 0.01794 \\
$\lambda_4$  &     1.11316  &     0.38252   &  \ 0.00436 \\
$\lambda_5$  &     0.91683  &     0.20340   &  \ 0.00000 \\
    \hline    
  \end{tabular}
\end{table}

\begin{figure*}
  \includegraphics[width=\textwidth]{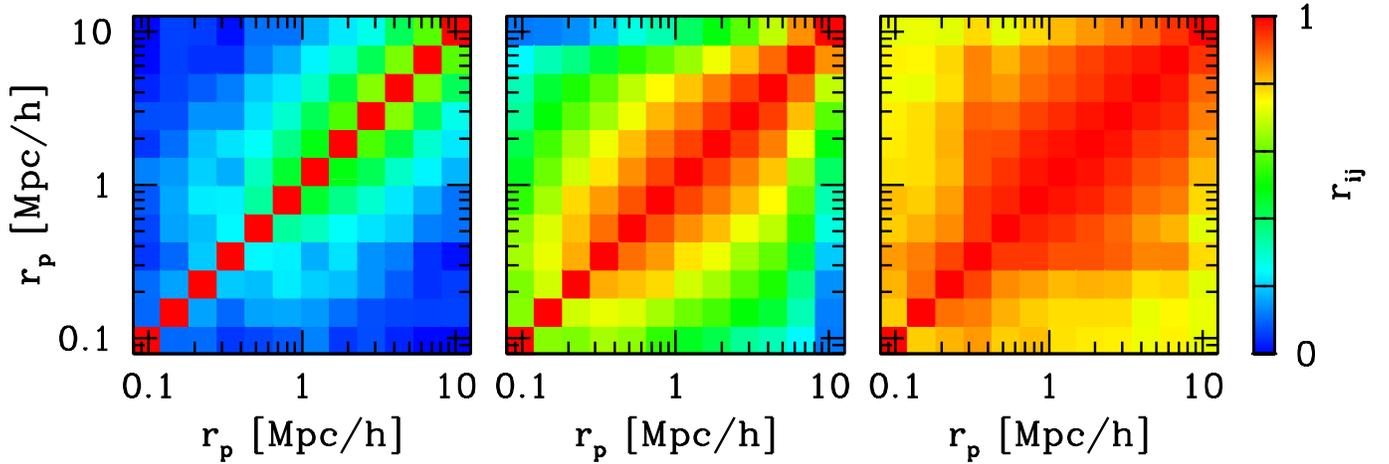}
  \caption{Mean of the 24 correlation matrix derived resampling the galaxies of eack KW24 mocks
    ({\it left}), or resampling 8 equal sub-volumes ({\it center}). These are compared
    to the correlation matrix derived directly from the KW24 mocks ({\it right}).
    The redshift range considered here is z=[0.5-0.8].
    The averaging over the 24 realisations of the 2 left matrices suppress
    the negative off-diagonal terms which are sometimes present for a given mock
    catalogues. Correlation coefficients are then color-coded from 0 to 1.
  } 
  \label{fig:correlation_matrix}
\end{figure*}

Each element of the the correlation matrix $r_{ij}$ is obtained
from the corresponding element of the covariance matrix $\sigma_{ij}$ as
$r_{ij}=\sigma_{ij}/\sqrt{\sigma_{ii} \sigma_{jj}}$. By definition,
the off-diagonal terms of the correlation matrix will then range between -1
and 1, indicating the degree of correlation between different scales
of the function $w_p(r_p)$.
Considering the redshift range z=[0.5-0.8], we show in Fig.~\ref{fig:correlation_matrix}
the mean of the 24 correlation matrices derived resampling the galaxies (left panel), or resampling
8 equal sub-volumes (center), 100 times each. These are compared to the correlation matrix
directly derived from the 24 mocks (right panel).
The first case shows a mainly diagonal correlation matrix where off-diagonal terms are mostly noise.
In the second case they instead decrease smoothly from 1 to 0 as a function of bin separation.
The matrix derived from the 24 mocks shows high correlation at all scales. 

In order to compare directly the properties of the correlation matrices
derived with the 3 methods, we diagonalize all of the $24+24+1$ matrices
by computing their principal components and the amplitudes of the
corresponding eigenvalues $\lambda_i$ ($i=1-12$).
Note that the sum of the eigenvalues of a correlation matrix 
is always equal to its dimension, i.e. 12 in our case. 
We report in Table~\ref{tab:eigencorr} the values of the five main eigenvalues
obtained with the 24 mocks (first column) compared to the averages
over the 24 mocks of those obtained with the 2 resampling methods.
The numbers show that the correlation matrix derived from the 24 mocks
contains essentially four principal components and is mostly dominated by one
of them. This indicates a strong correlation in the data.
The bootstrap matrices, instead, show more than 5 non-negligible components,
with the fifth one being of the same order of magnitude of the second
in the mock matrix. This implies a lower correlation.
We note, however, that volume resampling tends to
produce a matrix whose structure is closer to that of the mocks, with
1-2 dominant components.  This is another indication of how volume
bootstrapping, although also not reproducing perfectly the intrinsic
covariance properties of the sample, provides a better estimate of the
variance and correlation in the data with respect to
galaxy-galaxy bootstrap. 

These experiments are extended and further discussed in our parallel
accompanying papers, in particular by \citet{porciani2009}. The
bottom-line result of our extensive investigations is that volume
bootstrap, if enough resamplings are used, provides a sufficiently good
reconstruction of the intrinsic covariance matrix of the data set.
This is obtained at the expenses of a slightly less accurate
account of cosmic variance on large scale, with respect to what can
be obtained from the scatter among mock samples, where 
wavelengths longer than the survey size can be sampled.
However, we have shown (Fig.~\ref{fig:sigmeanbootmock}) that this
effect on scales $\sim 10$~h$^{-1}$~Mpc is limited to $\sim 20\%$.

\section{Dependence of galaxy clustering on luminosity}
\label{sec:results}

\subsection{Luminosity dependence at fixed redshift}
\begin{figure*}
 \includegraphics[width=\textwidth]{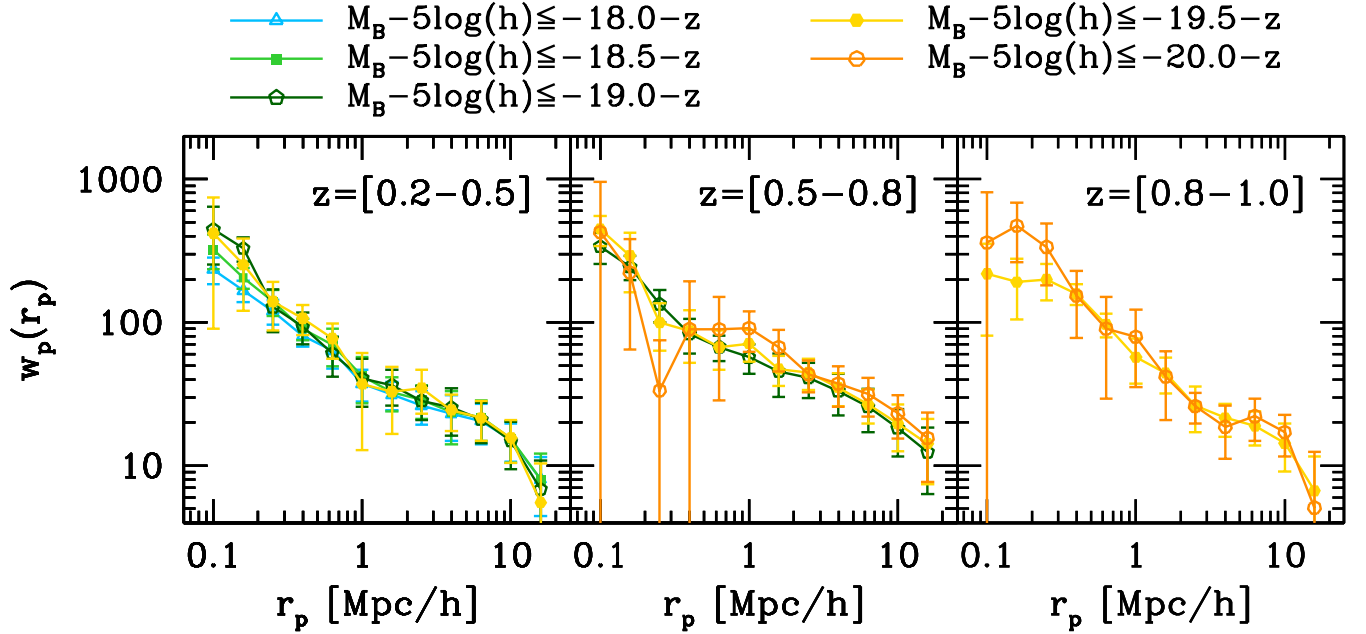}
 \caption{Projected correlation function \wprp\ measured 
   as a function of galaxy luminosity 
   within three redshift ranges. No significant dependence on
   luminosity is observed within the explored ranges.  Note the very
   flat slope of \wprp\ in the central redshift bin, compared to the
   two other slices.
 }
  \label{fig:wp_luminosity}
\end{figure*}
\begin{figure*}
  \includegraphics[width=\textwidth]{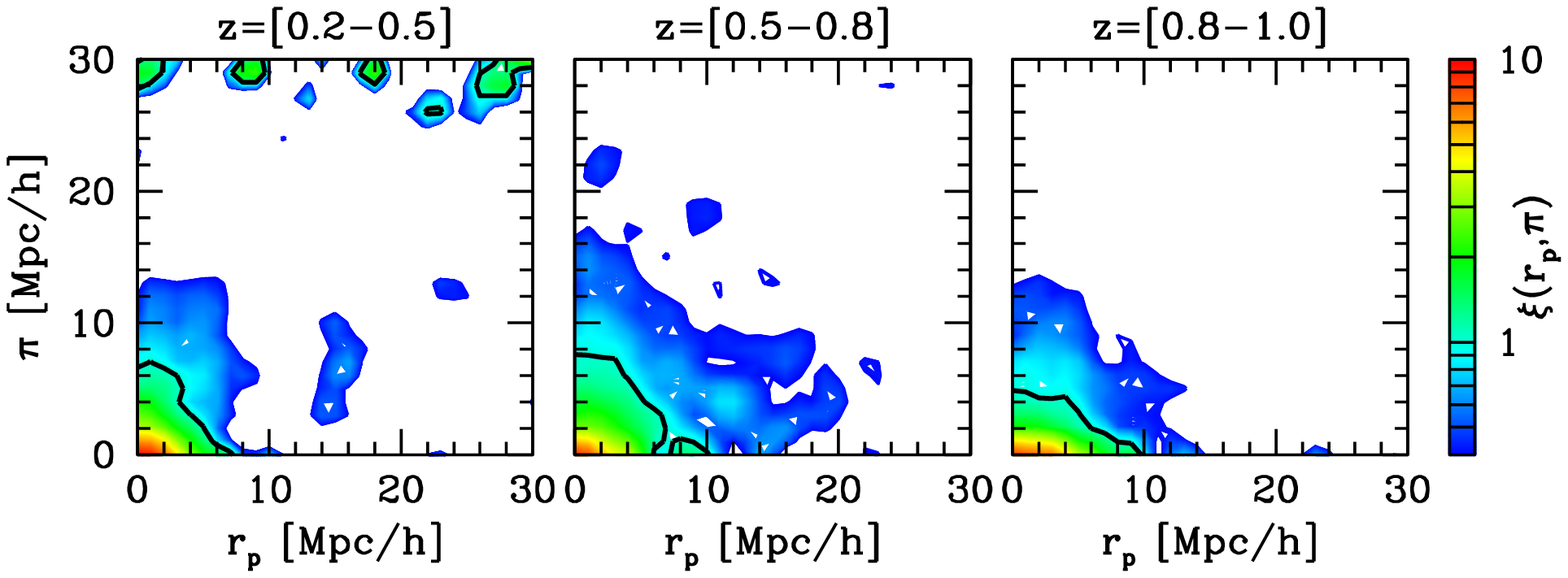}
  \caption{Iso-correlation contours of $\xi(r_p,\pi)$
    (here smoothed with a gaussian kernel) for galaxies brighter
    than $M_B-5log(h)\le -19.5 -z$, computed in 3 redshift ranges.
    The amplitude is  color-coded according to the side bar,
    while the black contour corresponds
    to $\xi(r_p,\pi)=1$. White values correspond to $\xi(r_p,\pi)<0.4.$} 
  \label{fig:xirppi_luminosity}
\end{figure*}

Figure~\ref{fig:wp_luminosity} shows the projected correlation function
\wprp\ estimated for our 9 luminosity selected sub-samples at
different redshifts.  Error bars correspond to the $1\sigma$
dispersion provided by 200 volume bootstrap resamplings, as extensively
discussed in Sect.~\ref{sec:xierrors}. 

We note that no clear dependence on luminosity is observed within any
of the three redshift ranges. Also, in the shape of \wprp\ there is some hint
for the usual ``shoulder'', i.e. a change of slope around 1~h$^{-1}$~Mpc,
but no clear separation between the expected 1-halo term on
small scales and 
the 2-halo component above this scale (see the Introduction for
definitions).  In particular in the intermediate redshift bin, all
sub-samples show a rather flat large-scale slope, with no
evidence for the usual breakdown above $\sim2$~h$^{-1}$~Mpc.

To try and understand the origin of the observed flat shape,
it is interesting to look directly at the contour plots of the
bi-dimensional redshift-space correlation function $\xi(r_p,\pi)$.
These are shown in 
Fig.~\ref{fig:xirppi_luminosity} for the three luminosity-selected
sub-samples L1.4, L2.2 and L3.1 (see Table~\ref{tab:lum} for definitions),
that include galaxies brighter than
$M_B-5log(h)\le -19.5 -z$. The three contour plots show some
interesting features. First, one clearly notices the much stronger
distortion along the line of sight $\pi$, in the central panel.  At
the same time, in the same redshift range a much more extended signal
is also observed along the perpendicular direction $r_p$.  It is tempting to
interpret both these effects as produced in some way by the two
dominating structures that we evidenced in
Fig.~\ref{fig:random_samp_radial_distrib}.  The 
excess signal along the line of sight is very plausibly due to the
distortions by 
``Fingers of God'', due to an anomalous number of virialized
systems (groups and clusters) within these structures.  
At the same time, the extension along $r_p$ is indicating that there
is also an excess of pairs perpendicularly to the line of sight, with
respect to an isotropic distribution.  In fact, we know
\citep{scoville2007b,guzzo2007} that the large-scale structure at
$z\simeq 0.73$ extends over a large part of the COSMOS area.
This evidently biases the
observed number of pairs along $r_p$, for simple geometrical reasons.
We cannot exclude that part of the large-scale compression observed in
$\xi(r_p,\pi)$ is also generated by an excess of galaxy infall onto
this structure, thus producing what is known as the Kaiser effect \citep{kaiser1987}.
This effect is proportional to the growth of structure
\citep[see e.g.,][for a recent direct estimate at similar redshift]{guzzo2008}
and can be extracted when the underlying clustering can be assumed to be isotropic.
In this case it is in practice
impossible to disentangle this dynamical distortion from the
geometrical anisotropy generated by having one dominating structure
elongated perpendicularly to the line of sight. 

The flatter shape in \wprp\ in Fig.~\ref{fig:wp_luminosity} is also consistent
with the overdense samples of \citet{abbas2007}, who noticed not only a higher amplitude
for the most overdense (10\% and 30\%) samples of mock and SDSS galaxies,
but also a flattening in the correlation function with respect to the full sample.
This is another line of evidence in favour of the hypothesis that the zCosmos field is
centered on an overdensity.

\begin{figure}
  \includegraphics[width=9cm]{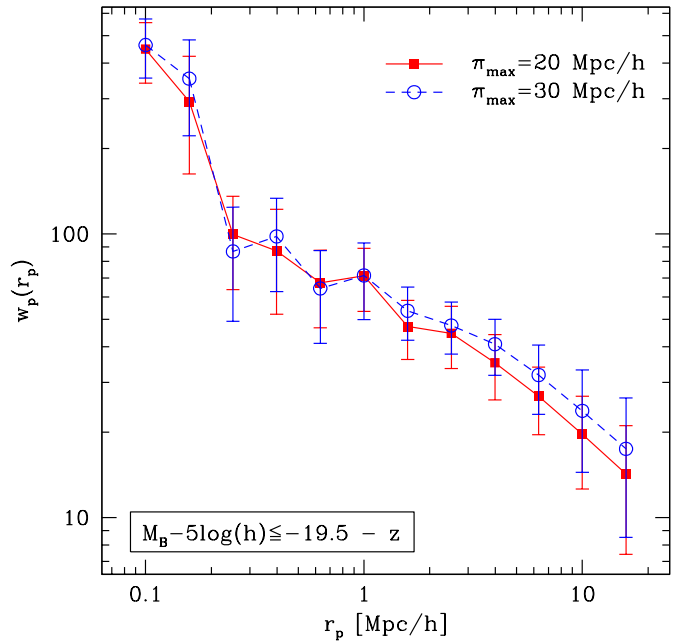}
  \caption{Sensitivity of the projected function \wprp\ to the upper
    integration limit $\pi_{max}$, for one luminosity-selected sample in the
    central redshift bin.}
  \label{fig:test_pimax}
\end{figure}

The plots of Fig.~\ref{fig:xirppi_luminosity} also show explicitly
the reasons for our choice  of
$\pi_{max}=20$~h$^{-1}$~Mpc as the upper integration limit in the
computation of \wprp: this value provides  a reasonable
compromise between including most of the signal and 
excluding the noisiest regions in the upper part the diagrams.
In the central redshift bin, however, some real clustering power
may still be present above this scale, for small $r_p$'s.
In Fig.~\ref{fig:test_pimax} we show directly how \wprp\ changes, when
$\pi_{max}$ is extended from 20 to 30~h$^{-1}$~Mpc. 
We see that, somewhat counter-intuitively, below 1~h$^{-1}$~Mpc
no extra amplitude is gained, while -- as indicated by the mock
experiments (see Sect.~\ref{sec:xi_technic}) -- the
scatter is increased.  Conversely, one can see the slight
scale-dependent bias on the amplitude at larger separations, which
gets up to $\sim 10\%$ at 15~h$^{-1}$~Mpc when increasing $\pi_{max}$.

%

\subsection{Redshift evolution at fixed (evolving) luminosity}

\begin{figure}
  \includegraphics[width=9cm]{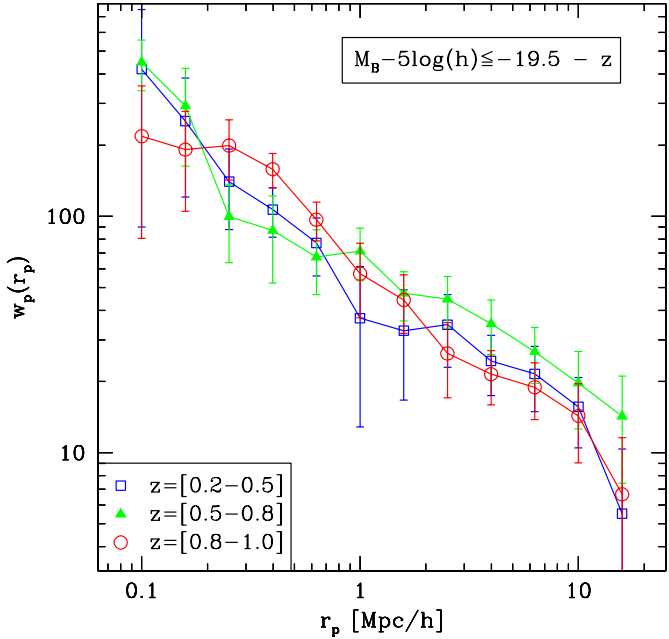}
  \caption{Evolution of the projected function \wprp\ of galaxies with
    $M_B-5log(h)\le -19.5 -z$ between redshift $z=0.2$ and $z=1$. }
  \label{fig:xilum_vs_z}
\end{figure}

In section \ref{sec:lumin_sel} we have discussed how our luminosity
selection has been devised as to account for the average evolution in
the luminosity of galaxies, assuming this to be the dominant effect in
modifying the mean density of objects at a given luminosity. Under
this assumption, it is then interesting to 
test how \wprp\ for galaxies within the same (de-evolved) luminosity
interval changes with redshift.  This is shown in
Fig.~\ref{fig:xilum_vs_z} for
$M_B-5log(h)\le -19.5 -z$, i.e. for the same three samples for which
$\xi(r_p,\pi)$ is plotted in Fig.~\ref{fig:xirppi_luminosity}. 
No coherent evolution of the
amplitude and shape of the projected correlation function is
observed among the three samples, characterized by mean
redshifts 0.37, 0.61 and 0.91. The three curves are consistent with each other
within the error bars, with -- as expected -- the
intermediate-redshift bin (triangles) showing a systematically
higher amplitude than the other two.
Again, this is easily intepreted as a local effect, resulting from the 
extreme large-scale clustering observed in this redshift range.  On the other
hand, the overall lack of apparent amplitude evolution of luminous
galaxies with redshift is consistent with previous results
from the VVDS-Deep \citep{pollo2006}, DEEP2
\citep{coil2006} and SDSS \citep{zehavi2005} surveys
for galaxies brighter than $\sim M^*$.  The only evolutionary effect
evidenced in particular by the VVDS data is a steepening of the
small-scale slope of \wprp\ (i.e. the 1-halo term) for high-redshift
luminous galaxies \citep{pollo2006}.
\citet{hjmcc2008} also show a lack of clustering amplitude
evolution for a large sample of 
luminous ($-22\le M_B-5log(h)\le-19$) galaxies based on accurate
photometric redshifts 
in the CFHTLS Deep fields; interestingly, they show that such
invariance is maintained also when splitting the sample into early-
and late-type galaxies.

\section{Dependence of galaxy clustering on stellar mass}

The relation of clustering properties to galaxy stellar masses is in
principle more informative and straightforward to interpret as stellar
mass is a more fundamental physical parameter than luminosity.  

\subsection{Mass dependence at fixed redshift}

\begin{figure*}
  \includegraphics[width=\textwidth]{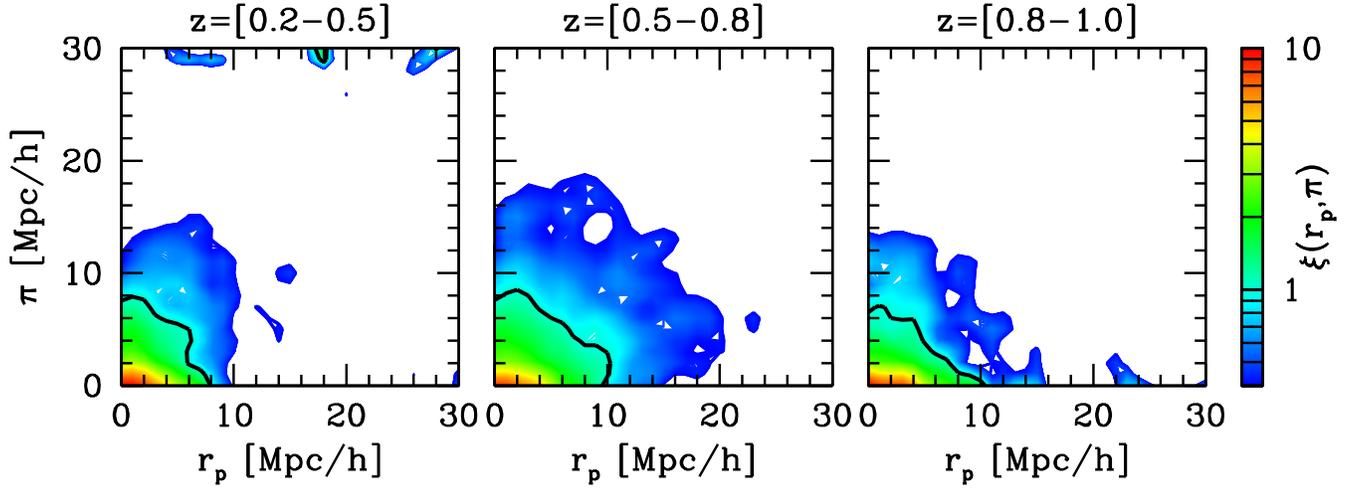}
  \caption{Example of full redshift-space correlation function
    $\xi(r_p,\pi)$ (here smoothed with a gaussian kernel)
    for the sub-sample containing galaxies more 
    massive than $10^{10}~h^{-2}~M_\odot$ computed in the same 3 redshift
    ranges as in Fig.~\ref{fig:xirppi_luminosity}.
    $\xi(r_p,\pi)$ is computed in cells of 1~h$^{-1}$~Mpc side in both
    $r_p$ and $\pi$ and the iso-correlation levels are color coded
    according to the side bar. The thick black contour corresponds to
    $\xi(r_p,\pi)=1$ and the white values to $\xi(r_p,\pi)<0.4$.
    Note how the central panel ($z=[0.5-0.8]$)
    shows extra power in both directions, perpendicular and parallel
    to the line-of-sight. The small-scale ``Finger-of-God'' effect is
    however less pronounced than in the case of the luminosity-selected
    sample of Fig.~\ref{fig:xirppi_luminosity}}  
  \label{fig:xirppi_mass}
\end{figure*}

Also in the case of stellar mass dependence, it is interesting to look
at the shape of the full correlation function
$\xi(r_p,\pi)$ in the three redshift ranges.
We show in
Fig.~\ref{fig:xirppi_mass}, the result obtained for the 3
samples M1.3, M2.3, M3.2 (see Table~\ref{tab:mass}), that include
galaxies more massive than $10^{10}~h^{-2}~M_\odot$.  Also in this
case the central panel is significantly different from the other two,
with $\xi(r_p,\pi)$ remaining positive out to much larger scales in
both $r_p$ and $\pi$ directions.  We note that the
small-scale stretching along $\pi$ seems to be less extended than that obtained
for the corresponding luminosity-selected sample, although it is hard
to say whether this difference is significative.  

\begin{figure*}
 \includegraphics[width=\textwidth]{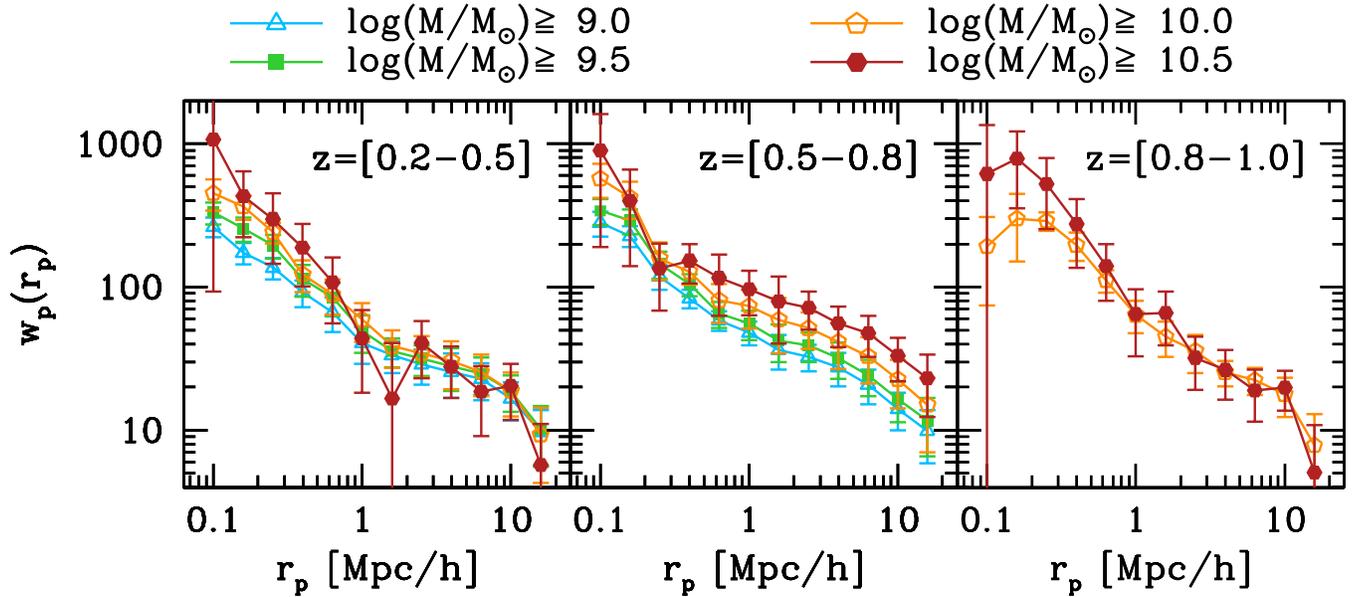}
 \caption{The projected correlation function \wprp\
   as a function of galaxy stellar masses in the zCOSMOS 10K sample,
   within three redshift ranges.   
 }
  \label{fig:wp_mass}
\end{figure*}

Figure \ref{fig:wp_mass} shows the projected correlation function
$w_p(r_p)$ of the 10 mass-selected samples. The plotted points are not
corrected for the residual stellar mass incompleteness
(see Sec.~\ref{sec:xi_incompleteness}). 
Errors are estimated as in the luminosity case using 200 
bootstrap resamplings of 8 equal sub-volumes of each dataset.  The
figure shows a weak mass dependence of clustering in the low- and
high-redshift bins, in particular at small separations.  At the same
time, a strong dependence at all separations is evident in the intermediate
redshift slice.  In this range the slope of \wprp\ remains extremely flat
out to the largest explored scales, even more strongly than in the
luminosity-selected cases.  Finally, we note also that in the low- and
high-redshift bins there is evidence for a steeper ``1-halo term''
contribution at $r_p<1$~h$^{-1}$~Mpc (with no clear indication for an
evolution in redshift of the transition scale to the 2-halo term).
Conversely, the central redshift range seems to be characterized by the
same, flat power-law shape down to 0.2~h$^{-1}$~Mpc, where a sudden
steepening is then observed. The slope below 0.2~h$^{-1}$~Mpc seems to
depend directly on the limiting mass, with more massive galaxies
showing a steeper correlation function.   In summary, no clear overall
trend can be evidenced among the three redshift ranges, with the central volume
again displaying peculiar clustering properties that apparently
dominate over any possible cosmological effect.
 
\subsection{Clustering evolution at fixed stellar mass}

\begin{figure}
  \includegraphics[width=9cm]{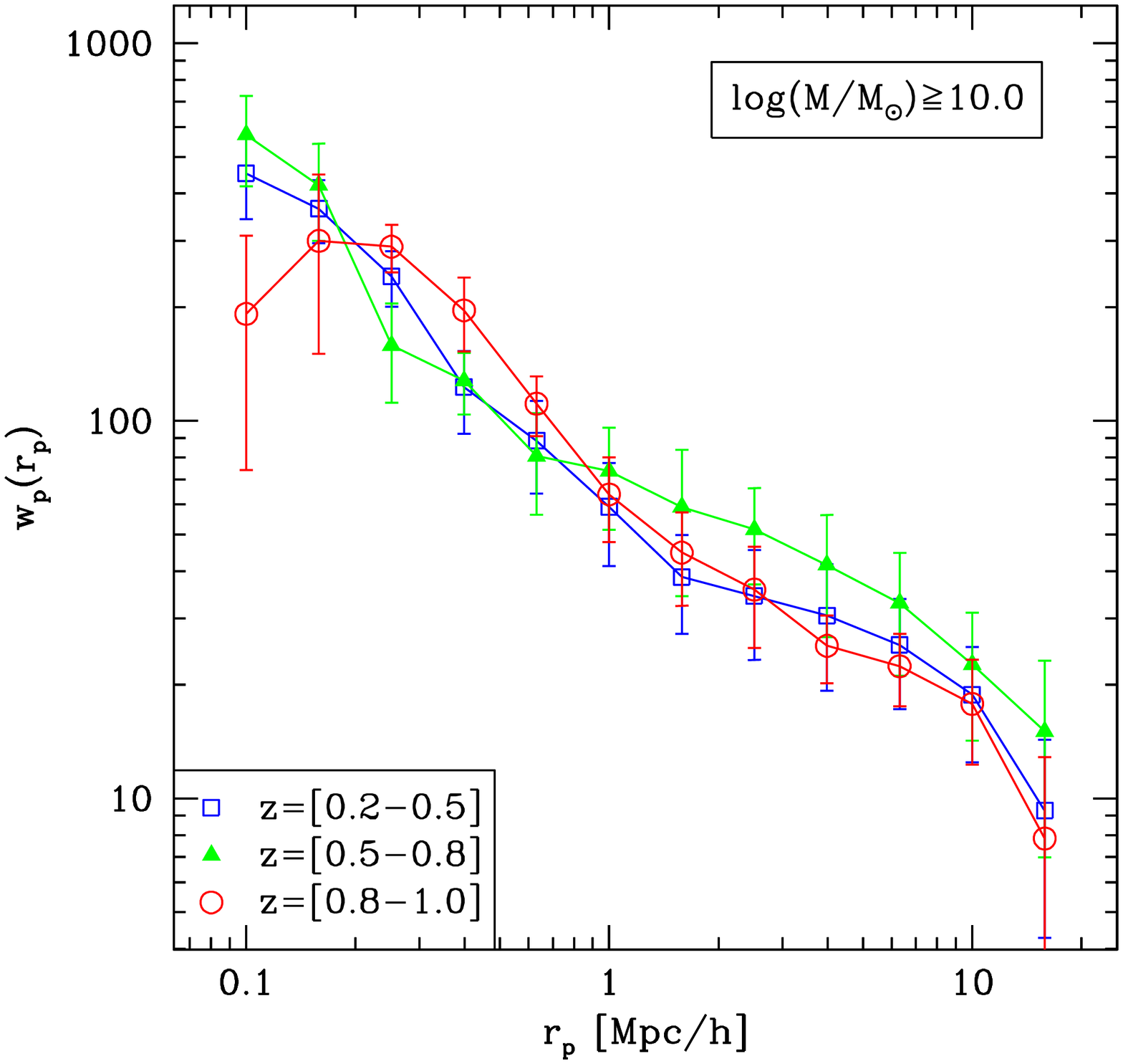}
  \caption{Evolution of the projected function \wprp\ of galaxies with
    $log(M/M_\odot)\ge10$  between redshift $z=0.2$ and $z=1$. }
  \label{fig:ximass_vs_z}
\end{figure}

It is also interesting to compare directly the evolution of \wprp\ with
redshift, when a specific class of stellar mass is selected.  As
mentioned earlier in this section, this is
particularly interesting, as in principle it does not require
accounting for strong galaxy evolutionary trends as in the case of
luminosity.  We are assuming here that stellar mass does not
significantly grow between $z\sim 1$ and $z\sim 0.2$, which we know is
only partially true.  A factor of $\sim 2$ growth in stellar mass is
in fact expected on average between $z=1$ and $z=0$, which
however would have little effect on the estimated correlation
function, if taken into account.
In Fig.~\ref{fig:ximass_vs_z} we
show \wprp\ computed for the same three samples with 
$log(M/M_\odot)\ge10$ of Fig.~\ref{fig:xirppi_mass}.  Similarly
to the luminosity samples, we do not
see any clear evolution with redshift of the amplitude and shape of
the projected correlation function.  The three curves are consistent
with each other within $1\sigma$.  

\section{Comparison with independent measurements and models} 
\label{sec:discussion}

\subsection{Redshift evolution of $w_p(r_p)$}

\begin{figure*}
  \includegraphics[width=\textwidth]{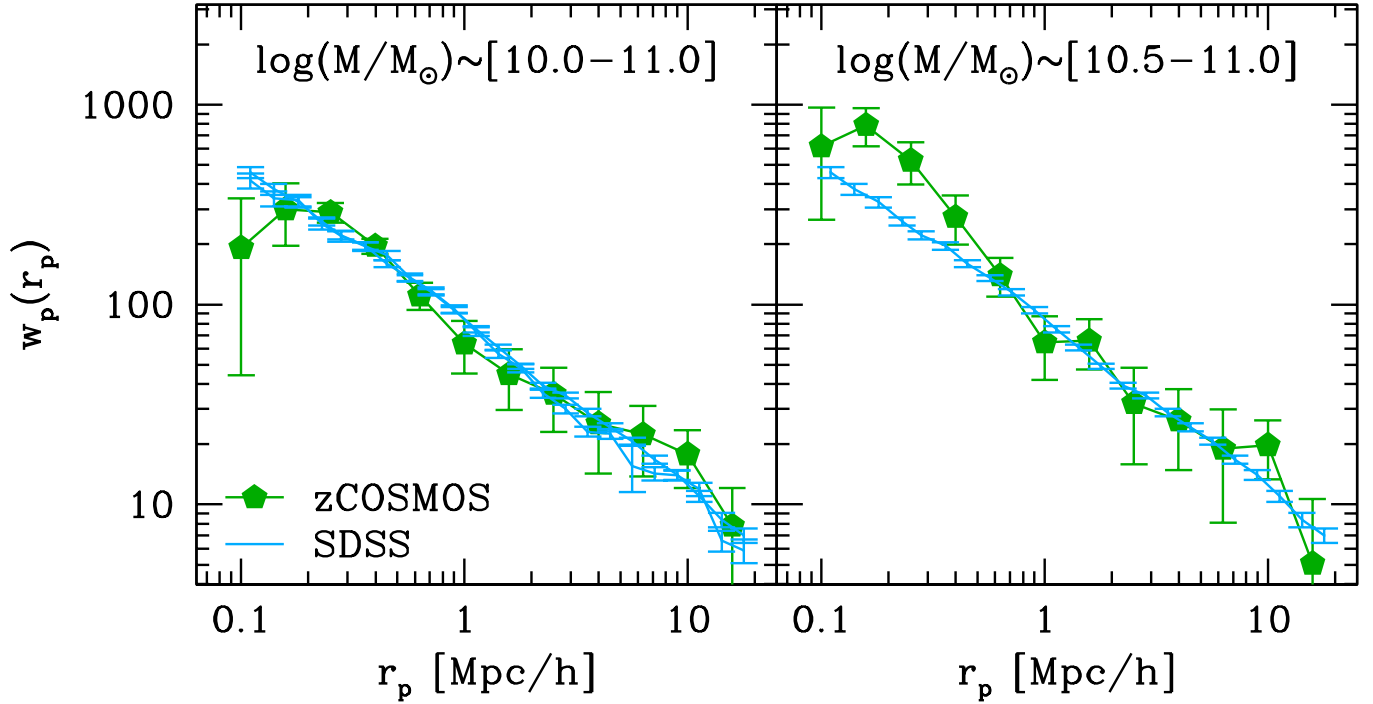}
  \caption{The measured projected function \wprp\ for
    galaxies at z=[0.8-1.0] with $log(M/M_\odot)\sim[10-11]$ ({\it left}) and 
    $log(M/M_\odot)\sim[10.5-11]$ ({\it right}) from the zCOSMOS
    survey (filled diamonds), compared to the $z\sim0.1$ estimate by
    the SDSS \citep{li2006} (blue curve).
  }
  \label{fig:ximass_zcosmos_sdss}
\end{figure*}

An accurate $z\sim 0$ reference measurement of $w_p(r_p)$ as a function of
stellar mass has been obtained by the SDSS \citep{li2006}.
\citet{meneux2008} do find evidence for evolution of the amplitude of
\wprp\ for galaxies less massive than $10^{10.5}~h^{-2}~M_\odot$,
when comparing this to the measurements from VVDS-Deep at $z\sim 0.9$.
The SDSS and zCOSMOS stellar masses were derived with the same
initial mass function \citep{chabrier2003} and normalised to h=1.
They are directly comparable.
The SDSS clustering measurements were obtained within differential
stellar mass ranges \citep{li2006}, while ours correspond to galaxies
more massive than a given threshold.
However, from Fig.~\ref{fig:lum_mass_samples} we see that the zCOSMOS
sample includes a very small number of galaxies more massives than
$10^{11}$~h$^{-2}~M_\odot$, due to the much smaller volume when
compared to the SDSS.  
Any of our mass-selected samples has therefore, in practice, an upper
bound at this value of mass.
This implies that we can coherently compare two of the 
SDSS measurements of $w_p(r_p)$
(for their galaxy samples with stellar masses in the ranges $[10.0-10.5]$ and $[10.5-11.0]$)
to those from our samples M3.1 and M3.2, that include galaxies more massive
than $10^{10}$ and $10^{10.5}$~h$^{-2}~M_\odot$ respectively, within
the redshift range z=[0.8-1.0].
This comparison (Fig.~\ref{fig:ximass_zcosmos_sdss}) does
not show a clear evolution with redshift. For both samples, the large-scale
amplitude of \wprp\ is virtually the same as in the local SDSS samples.
Considering simple evolution of structures,
this implies that the bias for galaxies more
massive than $10^{10}$~h$^{-2}~M_\odot$, has evolved significantly
between $z\sim 1$ and today, as to keep their apparent clustering
amplitude substantially unchanged.  
This implies in practice that the bias $b(z)$ must evolve in a
way such that $b(0)D(0)\simeq b(z)D(z)$, where $D(z)$ is the linear
growth rate of density fluctuations.  In the standard model, this
implies that at the approximate mean redshifts of our redshift bins,
$z=0.35, 0.75, 0.9$, the bias of massive galaxies must have been
respectively 1.2, 1.44 and 1.53 times its value at the current
epoch.
\citet{meneux2008} observed the same effect at $<z>\sim0.8$ in the
VVDS data but only for galaxies more massive than
$\sim10^{10.5}$~h$^{-2}~M_\odot$, with lower-mass objects showing
a weaker bias evolution.
A non evolution of the clustering of the most massive galaxies was also noticed in
the NDWFS \citep{brown2008} and 2SLAQ surveys \citep{wake2008}.

\subsection{Observed and predicted shape of $w_p(r_p)$ at $0.5<z<1$}

\begin{figure}
  \includegraphics[width=9cm]{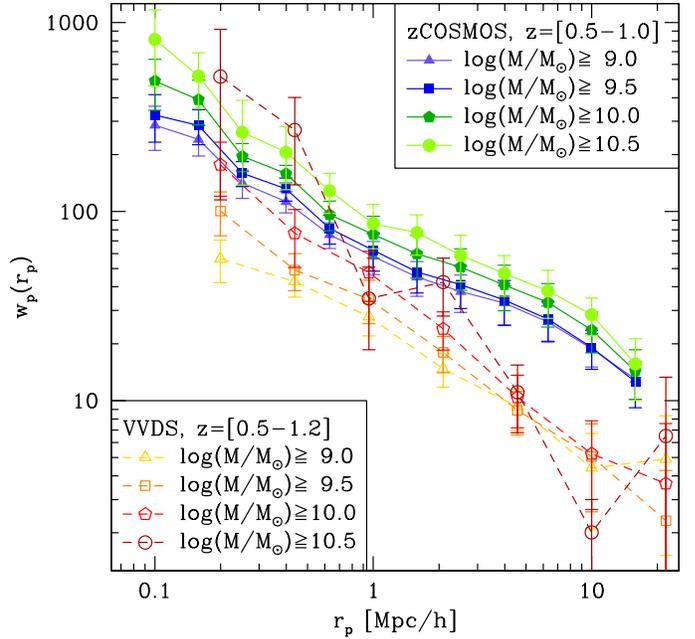}
  \caption{
    Direct comparison of the dependence of clustering on
    stellar mass in the VVDS-Deep and zCOSMOS samples, over a similar
    redshift range.
    This plot evidences the significant intrinsic
    difference between the two surveys, with the zCOSMOS sample showing a
    significantly larger clustering amplitude for all stellar masses.
    In addition, the much flatter shape of $w_p(r_p)$ 
    indicates the predominance of coherent structure perpendicularly to
    the line of sight, due to the ``walls'' discussed in the text.
  }
  \label{fig:wpmasszcosvvds}
\end{figure}

The only available measurement of clustering as a function of stellar
mass at redshifts comparable to those explored by our sample is that
from the VVDS-Deep survey \citep{meneux2008} at $0.5<z<1.2$.
VVDS-Deep goes 1.5 magnitude deeper (although over a smaller
area of $\sim 0.5$ deg$^2$), which allows the analysis to be extended
beyond $z=1$.  To provide a qualitative, yet meaningful comparison of
these two data sets, we can re-compute the correlation function for
the 10K data within the largest usable redshift range overlapping with
the VVDS interval, i.e.  [0.5-1.0].  We applied the same stellar mass
selection limits, keeping in mind the residual incompleteness that
will affect the highest redshift part of the sample.  The result is
shown in Fig.~\ref{fig:wpmasszcosvvds}, where the VVDS and zCOSMOS
mass-selected samples are directly compared.  The difference in shape
and amplitude in the \wprp\ derived from the two data sets is
rather striking.  The zCOSMOS points show in general a much
flatter relation than those from the VVDS.  The amplitudes for a given
mass selection also seem to be incompatible at several standard
deviations between the two samples, especially above 1~h$^{-1}$~Mpc. 

\begin{figure}
  \includegraphics[width=9cm]{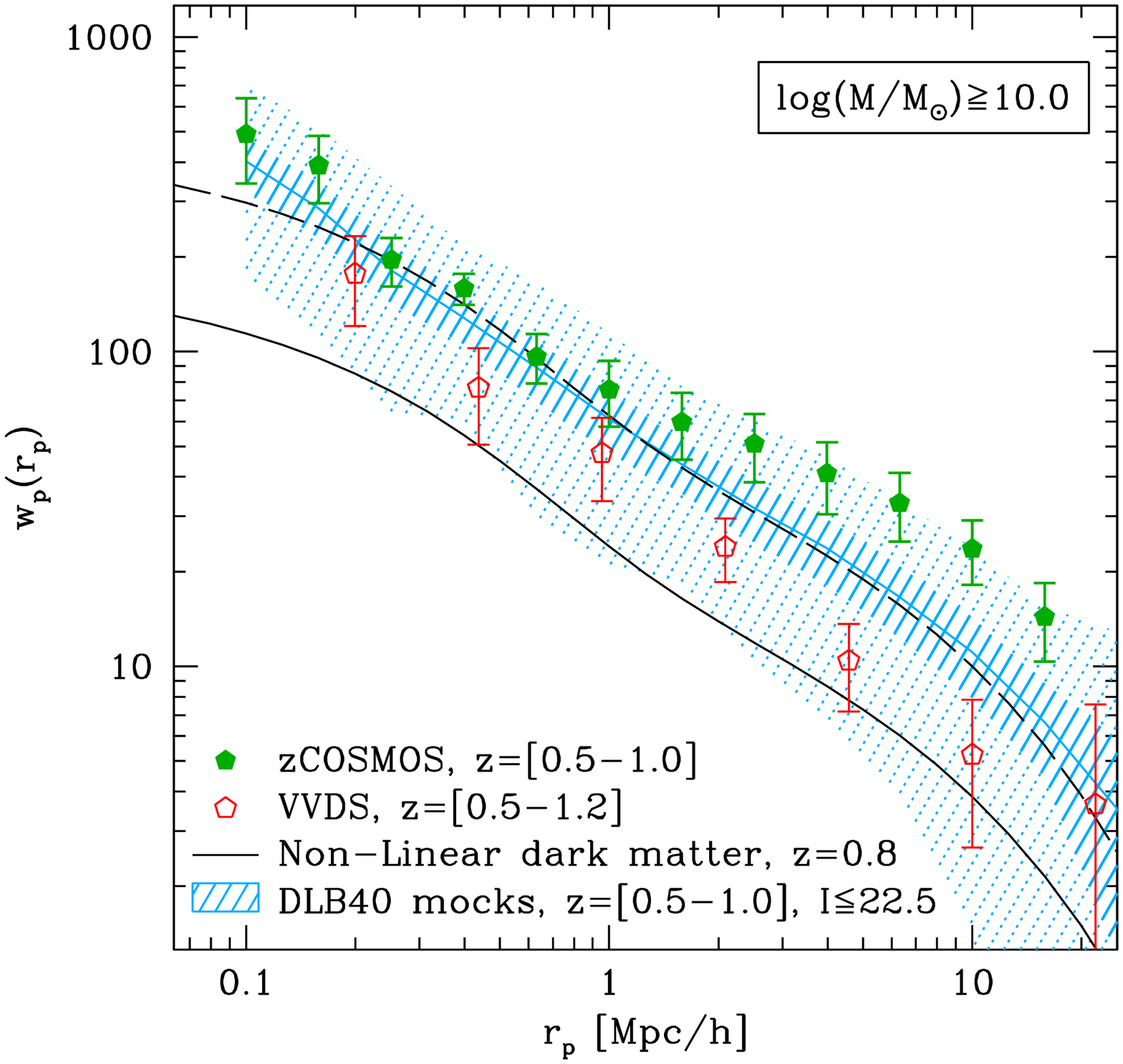}
  \caption{
    The zCOSMOS (solid circles) and VVDS-Deep (open
    circles) \wprp\ of galaxies
      with mass larger than $10^{10}~h^{-2}~M_\odot$ and $<z>\sim
      0.8$, compared to model 
      predictions.  These include the non-linear
      mass projected correlation function computed using {\it HALOFIT}
      \citep[lower solid line,][]{smith2003}, and, for reference, the
      corresponding \wprp\ for a population of halos with bias $b^2=2.6$
      (dashed curve).  The latter curve is a very good description of
      the full non-linear \wprp\ (light solid blue line),
      obtained averaging the forty DLB40 mocks from the Millennium run 
      after applying the same sampling, magnitude and mass selections
      of the 10K sample. The shaded areas (thick and thin shaded)
      give, respectively, the 
      $1\sigma$ and $3\sigma$ confidence corridors around the mean.  The
      large-scale zCOSMOS \wprp\ in this redshift range is thus
      marginally compatible with a rare, $\sim~3\sigma$ positive
      fluctuation in a standard $\Lambda$CDM Universe.}   
  \label{fig:wpmasszcosvvdsmodel}
\end{figure}

\subsection{Comparison to analytic and semi-analytic models}

It is at this point relevant to compare the available observations
from zCOSMOS and VVDS with model predictions in a standard $\Lambda$CDM
scenario.  We can do this in two ways.
We first used the {\sl HALOFIT} public code \citep{smith2003},
that uses the halo model to compute the expected non-linearly evolved power 
spectrum at $z=0.8$, that we take as a reasonable mean redshift for
the two samples (our conclusions would not differ at all if
predictions for $z=0.7$ or 0.9 were used).  
The corresponding projected function \wprp\ is then computed by
Fourier-transforming the power spectrum and projecting the resulting
real-space correlation function.  The result gives the expected \wprp\
of the mass density field at $z=0.8$.  Secondly, we can
compute the expectation value and the scatter expected in the same
redshift range ($0.5<z<1$) for \wprp\ using
the available semi-analytic mock surveys built from the Millennium
run.  To this end, we use the DLB40 mocks for which we have full
control over stellar masses, selecting simulated galaxies with
$I_{AB}\le 22.5$ and
$\log(M/M_\odot)\ge 10$, reproducing the sampling rate of the 10K data.
In Fig.~\ref{fig:wpmasszcosvvdsmodel} we plot both the {\sl HALOFIT}
prediction for the dark matter (lower solid line) and for
an arbitrarily biased population of halos with $b=\sqrt{2.6}\sim1.61$
(dashed line), together with the BDL40 mean \wprp\ (blue, lighter
solid line) and the corresponding $1\sigma$ and $3\sigma$ scatter corridor from
the 40 mock surveys (solid-
and dotted-dashed areas, respectively).  We note as a consistency
check the rather good agreement between the analytic {\sl HALOFIT}
result and the expectation value from the full n-body plus
semi-analytic simulation.  On these model predictions, we overplot the
corresponding zCOSMOS and VVDS estimates.  The zCOSMOS points agree 
rather well with the models (at better than 68\% confidence) in both
shape and amplitude on scales smaller than 1~h$^{-1}$~Mpc.  On larger
scales, however,  the observed \wprp\ would require
a strongly scale-dependent bias to be compatible with the model
predictions. Such a 
scale-dependence would also behave oppositely to what models and very
general considerations suggest, implying a bias which grows with
scale, rather than declines.  From the plot we see in fact that the
10K data are compatible with $b\simeq 1.6$ on small scales, but would
require $b\simeq 2.45$ on 10~h$^{-1}$~Mpc scales.  The shaded area
shows that this large-scale excess is marginally compatible with the
model predictions, representing a very strong positive fluctuation.
A few percent of volumes this size would show such a high
clustering amplitude (on these scales and for this kind of galaxies), in a
$\Lambda$CDM Universe.  

It is interesting to note, at the same time, how the VVDS measurements
lie on the opposite side of the distribution, at about
$1.5-2\sigma$ from the mean, but with a shape which is compatible with the
model prediction over the whole range (corresponding to a linear bias
$b\sim 1.2$). 
Comparison with \citet{abbas2007} also suggests that the zCOSMOS field is 
centered on an overdensity, whereas the VVDS field is a less significant underdensity.

These results show how a full HOD model fitting to the \wprp\
measured from the 10K data -- that we originally planned to include in
this paper -- would add no meaningful information to the
current analysis. Our first experiments with HOD models
based on the universal halo mass function
indicate that rather unrealistic sets of parameters are required
to reproduce the observed function.
An interesting possibility would be to use in such modeling a
halo mass function that depends on local environment
\citep[e.g.][]{abbas2005,abbas2006}, to take into account the
evidence that a large part of this sample is dominated by an
overdensity. We plan to explore this possibility using the
larger 20K zCOSMOS sample which is now nearly complete.

\section{Discussion}

Together with previous analyses \citep{hjmcc2007,kovac2009},
these results suggest that a significant fraction of the volume
of Universe bounded by the COSMOS field is indeed characterized
by particularly extreme density fluctuations.
We have seen how, in statistical terms, these seem to
lie at the $3\sigma$ limit of the distribution of amplitudes expected
in volumes of a few $10^6$~h$^{-3}$~Mpc$^3$. We should consider,
however, that these conclusions are drawn from measurements 
that are strongly affected by the angular distribution of 
structure. The \citet{hjmcc2007} result is based on the angular
correlation function, while here we have studied the projected
function \wprp. The latter, although making use of the redshift
information is in practice a clustering measure dominated by galaxy
pairs lying almost perpendicularly to the line of sight. The underlying
assumption when measuring \wprp\ is that the
geometrical distribution of structures within the sample being
analyzed is completely isotropic.  In other words, that there are
superclusters aligned along several directions, such that the only
remaining radial signal is produced by galaxy peculiar velocities.
The very reason of using \wprp\ is indeed to get rid of the
distortions introduced in the shape of $\xi(s)$ (the redshift-space,
angle-averaged correlation function) by galaxy motions. If
this is true, and only in this case, then \wprp\ is fully equivalent
to an integral over the real-space correlation function $\xi(r)$ and therefore
carries the same cosmological information.  However, if, as in the
case we have encountered here, there is one or more dominating
structures extending preferentially along one direction, then the use of
\wprp\ to infer cosmological information is inappropriate.

\begin{figure}
  \includegraphics[width=9cm]{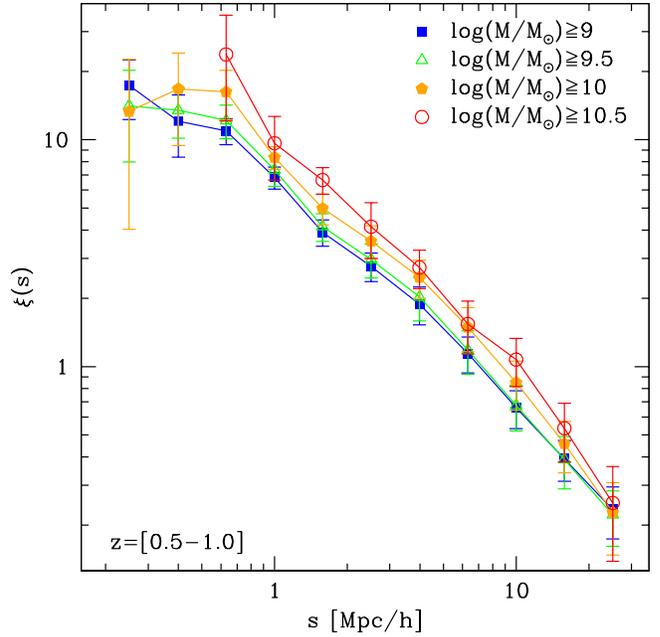}
  \caption{The redshift-space, angle-averaged correlation function
    $\xi(s)$ of the mass-selected samples in the
    redshift bin $[0.5,1]$.  A mild systematic mass dependence is
    visible.
  }
  \label{fig:xis-mass}
\end{figure}

One may thus wonder whether more robust cosmological
information could instead be inferred looking directly at the
simplest, angle-averaged redshift-space correlation function $\xi(s)$.
The expectation is that the average over all directions reduces the
weight of the excess pair counts produced by just a few
structures oriented along one preferential direction.  In such case
any analytic modelling (e.g. with HOD models) should also include an
appropriate model for the linear and non-linear redshift distortions
\citep{scoccimarro2004,tinker2007}. More simply, we can use the available
mock samples in redshift space to compute the non-linear
redshift-space $\xi(s)$ and its variance and compare
it to the data, as we did for \wprp. In Fig.~\ref{fig:xis-mass} we
first plot $\xi(s)$ for the 10K sample, computed for the usual four
mass ranges in the broad redshift range $0.5<z<1$.  The four
data sets show a smooth power-law behaviour, with some evidence for
a mass dependence of the clustering amplitude, in particular at the
upper mass limit.
The overall shape is well described by a rather flat power-law 
$\xi(s)\sim (s_0/s)^\gamma$, with slope $\gamma \sim 1$ and a
correlation length $s_0$ between 6 and $\sim10$~h$^{-1}$~Mpc.
These values for the shape and amplitude of $\xi(s)$ are similar to
those measured for luminous red
galaxies at $z=0.55$ in the 2SLAQ survey (see Fig.~7
in \citet{ross2007}) and for luminous early-type galaxies in
the 2dFGRS \citep{norberg2002}.  This is consistent with the most
massive objects in the 10K sample being predominantly red, early-type
galaxies which show moderate or no evolution in the overall clustering
amplitude with redshift.  

\begin{figure}
  \includegraphics[width=9cm]{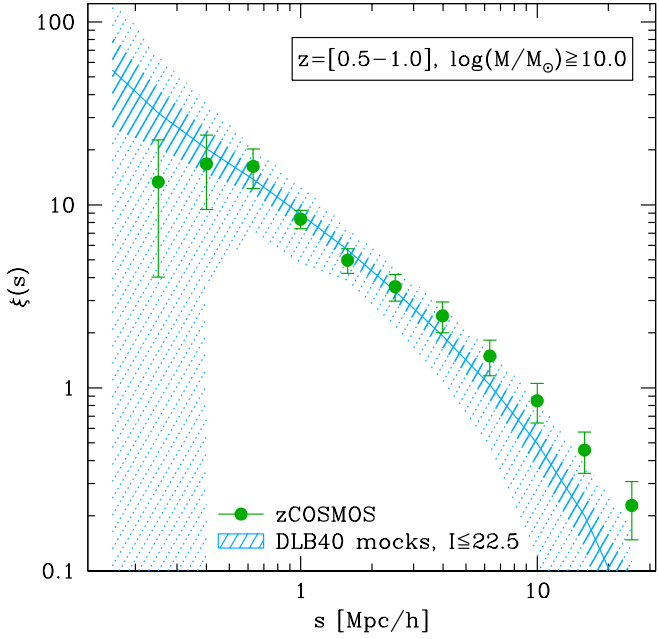}
  \caption{Comparison of the redshift-space correlation function
    for galaxies with $\log(M/M_\odot)\ge 10$ at $0.5<z<1$ in the
    zCOSMOS sample (filled circles), with the model predictions from
    the BDL40 mock samples. The solid line gives the average of the 40
    mocks, with the dashed areas corresponding to the $1\sigma$ and $3\sigma$ confidence
    error corridor. Similarly to what we found for \wprp,
    the agreement between the models and the zCOSMOS
    measurement at 1~h$^{-1}$~Mpc is remarkable.
    Nevertheless, the large-scale shape of the zCOSMOS $\xi(s)$
    is at $\sim~3\sigma$ the mean amplitude of the mock catalogues.
  }
  \label{fig:xis-models}
\end{figure}

In Fig.~\ref{fig:xis-models}, instead, we compare $\xi(s)$ of our
``reference'' sample with $\log(M/M_\odot)\ge 10$, with the mean and
scatter (at $1\sigma$ and $3\sigma$ confidence, respectively) of the
similarly-selected set of BDL40 mocks. Despite the angular average,
we note a behaviour which is similar to that observed in \wprp\, although
now the agreement extends to slightly larger scales. The observed
clustering is compatible with the predictions of the standard model (to
better than the 68\% level) on scales smaller than $\sim~2$~h$^{-1}$~Mpc.
On larger scales, also $\xi(s)$ shows excess power
with respect to the models, which places the zCOSMOS volume at the
upper $3\sigma$ limit of the statistical distribution obtained from
the mocks.  This exercise shows that even after angle-averaging our
clustering estimator, the amount of structure present in this
specific volume of the Universe remains outstanding in comparison to
the model expectations.  The conclusion can only be that either we
have been very unlucky in the selection of the COSMOS field and picked
up a fluctuation which has a probability of $\sim$~1\% to be found
in such a volume, or fluctuations with this amplitude are in reality more
common than what the standard cosmology predicts.

\section{Summary}
\label{sec:summary}
We have used the 10K zCOSMOS spectroscopic sample to study galaxy clustering
as a function of galaxy luminosity and stellar mass, in the range of redshift [0.2,1]. 

To this end, we built luminosity and mass-selected samples from the 10K
catalogue sampling three separate redshift ranges. We used mock
catalogues to quantify the effect of stellar 
mass incompleteness on the measured clustering, as a function of
redshift.  
We carefully checked our covariance and error estimate techniques,
comparing the performances of methods based on the scatter in the
mocks and on bootstrapping schemes. We adopted the latter, based on
200 resamplings of 8 sub-volumes of the survey, as the most appropriate
description of of the covariance properties of the data.

By measuring the redshift-space correlation functions $\xi(s)$ and
$\xi(r_p,\pi)$ and the projected function $w_p(r_p)$ for these
sub-samples, we found the following results.  
\begin{itemize}
\item Surprisingly, we do not see any clear dependence on luminosity
  of the correlation function at all redshifts.  This is at odd with
  results in the local Universe by the 2dFGRS and with 
  mesurements at similar redshift by the VVDS and DEEP2 surveys, that
  found a significant steepening of \wprp\ with luminosity.

\item We find a mildly more evident (although not striking)
  dependence of \wprp\ on stellar mass, especially on small scales.
  The central redshift bin 
  ($0.5<z<0.8$), displays in general a more evident effect, with a
  very flat shape of \wprp\ on scales
  $r_p=[1-10]$~h$^{-1}$~Mpc.   The overall shape of the corresponding
  map of $\xi(r_p,\pi)$ shows strong distortions that we interpret as
  the effect of dominant structure extending preferentially
  perpendicularly to the line of sight. 

\item From comparison to the SDSS measurements at $z\sim 0$, we do not
  see any significant evolution with redshift of the 
  amplitude of clustering for bright and/or massive galaxies. Together
  with previous results from VVDS, this is consistent with a more
  rapid evolution of the linear bias for the most massive objects, with
  respect to the general population.  In the zCOSMOS sample this
  invariance in the clustering amplitude between $z\sim 1$ and $z\sim
  0$ seems to remain valid down to smaller masses than in the VVDS, an
  effect easily explained by the overall larger clustering amplitude
  observed in general for $z>0.5$ in this sample.  This is evidenced
  by a much flatter shape (higher amplitude) of \wprp\ of zCOSMOS
  galaxies with respect to VVDS galaxies, when selected with the same
  criteria. 

\item This particularly high level of structure is confirmed by
  comparison of the measured \wprp\ and $\xi(s)$ at $0.5<z<1$ with model
  predictions, concentrating on the sample with $\log(M/M_\odot)\ge
  10$.  On scales smaller than $\sim 1-2$ h$^{-1}$ Mpc, the
  observations agree very well with the model expectation values in the standard
  $\Lambda$CDM scenario for a linar bias $b \sim 1.6$.  On these
  scales, the measured values are compatible to better than 68\% with
  the BDL40 mocks.  On larger scales, however, the observed clustering
  amplitude is reproduced in only a few percent of the mocks. In
  other words, if the shape of the power spectrum is that of
  $\Lambda$CDM and the bias has no ``innatural'' scale-dependence,
  COSMOS has picked up a volume of the Universe which is rare,
  $2-3\sigma$ positive fluctuation.   This conclusion is corroborated
  also by comparison with the VVDS measurements, which on the other
  hand lie on the lower side of the distribution, at about $1.5-2\sigma$.

\end{itemize}


\begin{acknowledgements}
  We thank the anonymous referee for a detailed review of the manuscript that helped to improve the paper.
  We thank G.~De~Lucia, J.~Blaizot, S.~Phleps and A.~S\`anchez for their thorough comments
  on an early version of the manuscript.
  This work was supported by grant ASI/COFIS/WP3110 I/026/07/0.

\end{acknowledgements}

\end{document}